\documentclass[12pt,showpacs]{revtex4}

\usepackage[english]{babel}
\usepackage{amsbsy}
\usepackage{amstext}
\usepackage{pst-node}
\usepackage{graphicx}
\usepackage{amssymb}
\usepackage{epstopdf}

\newcommand{\be}{\begin{eqnarray}}
\newcommand{\ee}{\end{eqnarray}}
\newcommand{\ba}{\begin{array}}
\newcommand{\ea}{\end{array}}

\newcommand{\paq}[1]{\left[#1\right]}

\newcommand{\ds}{\displaystyle}

\begin{document}

\title{Coherent detection method of gravitational wave bursts
for spherical antennas}
\author{Stefano Foffa and Riccardo Sturani}
\affiliation{D\'epartment de Physique Th\'eorique, Universit\'e de Gen\`eve, 
Geneva, Switzerland\\
Stefano.Foffa, Riccardo.Sturani@unige.ch} 

\begin{abstract}
We provide a comprehensive theoretical framework and a quantitative test of 
the method we recently proposed for processing data from a spherical 
detector with five or six transducers. 
Our algorithm is a trigger event generator performing a coherent analysis 
of the sphere channels.
In order to test our pipeline we first built a detailed numerical model of the 
detector, including deviations from the ideal case such as quadrupole modes 
splitting, and non-identical transducer readout chains.
This model, coupled with a Gaussian noise generator, has then been used to 
produce six time series, corresponding to the outputs of the six
transducers attached to the sphere.
We finally injected gravitational wave burst signals into the data stream,
as well as bursts of non-gravitational origin in order to mimic the 
presence of non-Gaussian noise, and then processed the mock data.
We report quantitative results for the detection efficiency versus false alarm 
rate and for the affordability of the reconstruction of the direction of 
arrival. In particular, the combination of the two direction reconstruction 
methods can reduce by a factor of 10 the number false alarms due
to the non-Gaussian noise. 
\end{abstract}

\maketitle

\section{Introduction}
Started more than forty years ago with the pioneering work of Weber 
\cite{Weber}, the rush for direct gravitational wave (GW) detection have 
nowadays reached a crucial phase. On one side, the era of large
interferometers (LIGO, VIRGO, GEO, TAMA) \cite{intf_web} 
has begun and the next, so-called ``advanced'' versions of the interferometers
are currently considered as the most likely candidate to make the first 
detection (although uncertainties in stellar population estimates still do not 
allow to keep this goal for granted, see \cite{deFreitasPacheco:2006bh}).

On the other side, resonant bars \cite{bar_web} like the INFN ones 
(EXPLORER, NAUTILUS, AURIGA) or ALLEGRO seem to have exhausted their original 
leading role after several years of impressive performances (see for instance 
\cite{Astone:2006ut} for some reports on the most recent runs) which made them 
by far the best instruments for a long time.

This does not mean at all that resonant detectors are going to be out of 
business, as resonant bars are currently running as a 
network \cite{Astone:2007vu} and joint analyses with interferometers
have already taken place \cite{Collaboration:2007paa} (and are foreseen in the 
future). Moreover high-performance, new generation antennas (like large
spheres \cite{Fafone:2006nr} or dual detectors 
\cite{Bonaldi:2003at}) have already been conceived. In this framework, the two 
small spherical detectors in Leiden \cite{Gottardi:2007zn} and San Paulo 
\cite{Aguiar:2006va}, that are already in the commissioning phase, can 
rightfully be considered as the first specimens of this new generation.

What makes a sphere really different from bars and interferometers is the fact 
of being a multichannel detector.
This distinguishing feature not only determines its isotropic sensitivity and
its capability to reconstruct the GW direction, but also opens the way
to new opportunities, and issues, from the data analysis point of view.
After the seminal work of Wagoner and Paik \cite{Wagoner}, where the basic 
features of the detector have been pointed out, more and more detailed sphere 
models and configurations have been studied by several authors 
\cite{Zhou:1995en,Coccia:1995yi,Lobo:1996ae,Stevenson:1996rw,Merkowitz:1997qc,Merkowitz:1997qs,Lobo:2000hy,Gasparini:2006vb,Gottardi:2006gn}.
In particular, various solutions have been proposed to the problems of the 
ideal transducers configuration \cite{Lobo:2000hy}, parameter reconstruction 
\cite{Merkowitz:1997qs,Zhou:1995en}, and multidimensional data analysis 
\cite{Stevenson:1996rw}.

The analysis method that we have set up and studied in the present work is the 
result of a synthesis and a refinement of some of these contributions, and is 
intended to represent the core of the pipeline that we are building for the 
miniGRAIL detector. Some preliminary results have already been proposed in
\cite{Foffa:2008pm}, here we expose the theoretical foundations and assess
quantitatively the solidity of the method as an event trigger-generator.

Particular care has been dedicated to all those
concrete  details that enter the game as ``theoretical'' data analysis is 
turned into a real, working device: these include, {\it in primis}, attention 
to the problem of false detections and keeping under control the computation 
time required by the pipeline.  

The first part of the paper is devoted to building an accurate detector model 
and to the generation of a set of simulated data, expanding the treatment in 
\cite{Foffa:2008pm}. We consider a real, imperfect sphere with six read-out
electromechanical amplificators of its surface oscillations. Our method 
works also with five outputs, while using a smaller number of transducers 
significantly degrade the performances of the detector, even if four 
transducers are {\it in principle} sufficient to fully reconstruct the GW 
tensor.

After having written the basic equations, we work out the 
spectrum of each transducer's output and we generate simulated data starting 
from the elementary noises which enter the detector at different 
levels. Then we discuss the problem of extracting the GW parameters
out of the data.

In the second part we describe the method to generate triggers of events, 
based on the coherent WaveBurst algorithm, the event trigger generator used by
the LIGO-VIRGO burst data analysis group. We adapted it to a multimode 
detector, where different channels have correlated 
noise, differently from the LIGO-VIRGO situation where the detectors are not 
co-located.
The availability of several channels allows a determinaton of the arrival
direction of the burst through a standard \emph{likelihood} analysis method.
Such a determination is then validated by a consistency check with the 
geometrical method exposed in paragraph \ref{subsdete}, named 
\emph{determinant} method, which provides an independent indication of 
the arrival direction.

In the last section the efficiency of the method for four different amplitudes 
of injected bursts is shown. We tested the method against injections of both 
\emph{GW} and \emph{non-GW} signals, where for non-GW signals the injection 
amplitude is shared randomly the five quadrupolar channels.

While each of the two above-mentioned method alone is not able to distinguish
between GW and non-GW excitations, we find that the combination ot the two 
allows to reduce the false alarm rate by roughly one order of magnitude, while 
keeping a good detection efficiency, and provides a good determination of the 
arrival direction of the signal. The multi-mode coherent analysis is thus able 
to compute all the relevant parameters of a GW and to reduce the false alarm 
rate.

\section{Model of the detector}

\subsection{Equations for modes, transducers and readout currents}
The sphere is well modeled as a set of coupled oscillators,
which describe the dynamics of the relevant sphere vibrational modes,
of the transducers, and of the electrical circuit that are at the core of the 
readout devices.
As the relevant equations have been already discussed in detail by several
authors  (see for instance
\cite{Merkowitz:1997qc,Gottardi:2006gn,Maggiore:1900zz}),
we jump to the mathematical core of the problem, skipping 
introductory material and definitions that can be found in the literature.
These equations hold for any number of sphere modes and transducers,
although in the following subsections we will specify our choices for such 
numbers. 
For the sake of clarity and completeness, the definitions and values of all the
quantities appearing below are summarized in fig.\ref{readout} and in Appendix 
A.

\begin{figure}
\includegraphics[width=1.\linewidth]{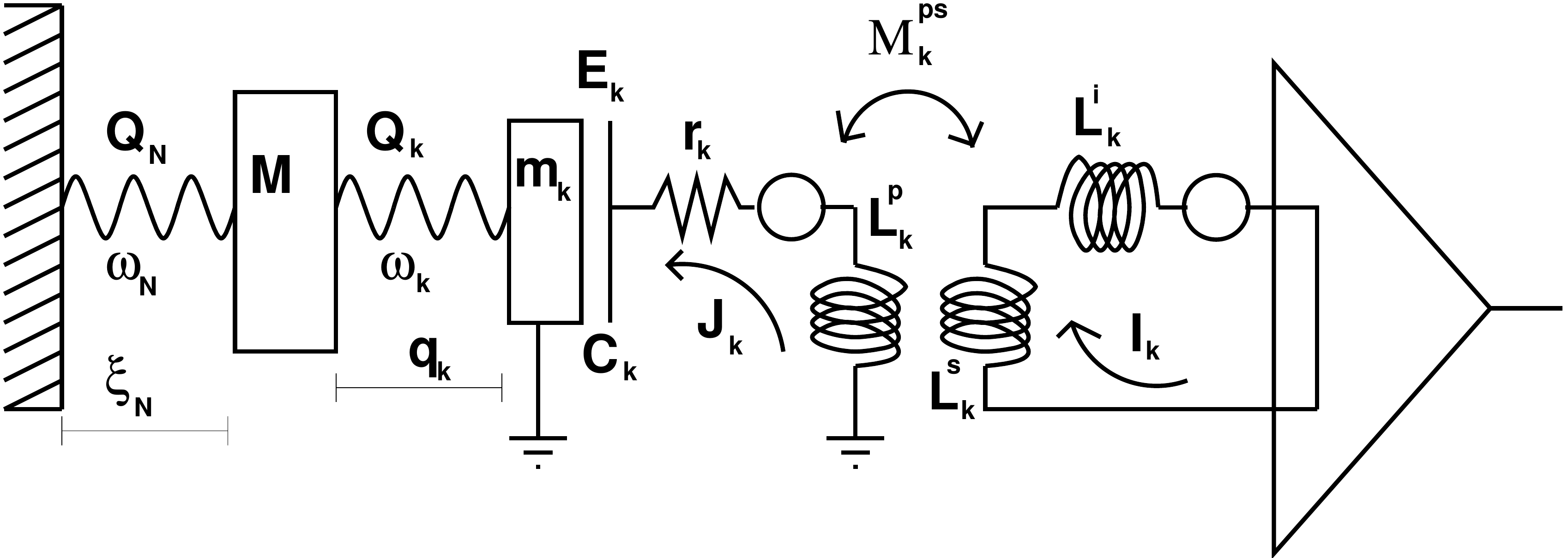}
\caption{The readout scheme, from \cite{Foffa:2008pm}.}\label{readout}
\end{figure}

In Fourier space, the amplitudes $\xi_N$ of the various radial modes of the 
sphere obey to the following equations
\be
\label{modeline}    
\ba{rl}
\ds
\!\!\!\!\!\!\!\!\!\!\!\!
\left(\omega_N^2-\omega^2+i\frac{\omega_N\omega}{Q_N}\right)\xi_N-\alpha_N 
\sum_k B_{Nk}
\left[\left(\omega_k^2+i\frac{\omega_k\omega}{Q_k}\right)\frac{m_k}{M} q_k-
i\frac{E_k}{M \omega}J_k\right]&=\\
\ds
\!\!\!\!\!\!\!\!\!\!\!\!
f_N-\alpha_N \sum_k B_{Nk} \frac{m_k}{M}f^t_k& .
\ea
\ee
which relate their dynamics to that of the transducer displacements $q_k$ and 
of the primary currents $J_k$ through the position matrix 
$B_{Nk}\equiv Y_N(\theta_k,\phi_k)$\footnote{We use here the real spherical harmonics, whose exact definition can be found in \cite{Zhou:1995en}.}.
The equations for the transducer displacements are:
\begin{eqnarray}\label{trasdline}    
-\omega^2 \sum_N B_{kN} \alpha_N\xi_N+
\left(\omega_k^2-\omega^2+i\frac{\omega_k\omega}{Q_k}\right)q_k
-i\frac{E_k}{m_k \omega} J_k=f^t_k\, ,
\end{eqnarray}
while the readout circuit can be described through the equations driving the 
primary currents and the secondary ones, $I_k$, which are proportional to
the actual outputs of the detector: 
\begin{eqnarray}\label{primline}    
E_k q_k +\left[r_k+i\left(\omega L^p_k - \frac{1}{\omega C_k}\right)\right] 
J_k -i\omega M^{ps} _k (I_k-f^i_k)=f^p_k\, ,
\end{eqnarray}
\begin{eqnarray}\label{lastline}    
-i\omega M^{ps}_k J_k + i\omega \left(L^s_k+L^i_k\right) (I_k-f^i_k) =f^s_k\, .
\end{eqnarray}
The $f$'s appearing above are the stochastic forces related to the various 
dissipative components of the detector\footnote{In subsection 
\ref{subs_transfh}, the $f_N$'s will also be used to describe the
deterministic interaction of the GW with the sphere.}; their action can be 
described through the corresponding noise spectral densities:
\begin{eqnarray}\label{noisedef}    
<f_x(\omega) f^*_y(\omega')>\equiv S_{xy}(\omega)\delta(\omega-\omega')\, .
\end{eqnarray}
The non-vanishing components of the noise spectral density matrix are:
\begin{eqnarray}\label{noisemt}    
S_{NN'}=2 k_B T \frac{\omega_N}{M Q_N} \delta_{NN'}\,,\quad\!
S^t_{kk'}=2 k_B T \frac{\omega_k}{Q_k m_k} \delta_{kk'}\,,\quad\!
S^p_{kk'}=2 k_B T  r_k \delta_{kk'}\,,
\end{eqnarray}
and, according to the Clarke model \cite{Clarke} for the SQUID,
\begin{eqnarray}
\label{noiseSQ}    
S^s_{kk'}&=&16\frac{k_B T}{R^{sh}_k}\left(\frac{L^{SQ}_k}{M^{SQ}_k}\right)^2\delta_{kk'}\,,\nonumber\\
S^i_{kk'}(\omega)&=&11\omega^2\frac{k_B T}{R^{sh}_k}(M^{SQ}_k)^2
\delta_{kk'}\,,\\
S^{s,i}_{kk'}(\omega)&=&-12 i\omega\frac{k_B T}{R^{sh}_k}L^{SQ}_k\delta_{kk'}
\,,
\nonumber
\end{eqnarray}
where $S_{NN'},S_{kk'}^t,S^p_{kk'},S^s_{kk'},S_{kk'}^i,S^{s,i}_{kk'}$ are the 
spectral densities of, respectivley: the forces acting on the the radial mode 
displacements, on the transducers, on the primary currents, on the secondary 
currents, the spectral densities of the secondary current noises and the 
spectral density of the secondary current noise forces times the secondary 
current noise. 

\subsection{Output current determination and simulation}
\label{subs_simul}
\subsubsection{Specification of the system}
~\\
By using the following matrix notation
\begin{eqnarray}
\ds {\cal Z}\equiv\nonumber\\
\ds\!\!\!\!\!\!\!\!\!\!\!\!\!\!\!\!\!\!\!\!\!\!\!\!\!\!\!\!\!\!\!\!\!
\left(\begin{array}{c|c|c|c}{\cal D}\left[\Omega_N\right]
& - {\cal D}\left[\alpha_N\right] \cdot {\cal B}_{Nk} \cdot{\cal D}
\left[ \tilde{\Omega}_k \right]
&{\cal D}\left[\alpha_N\right]\cdot {\cal B}_{Nk} \cdot {\cal D}
\left[\varepsilon_k/m_k\right]& 0 \\
\hline - \omega^2 {\cal B}^T_{kN}  \cdot{\cal D}\left[\alpha_N\right] & 
{\cal D}\left[\Omega_k\right]
& - {\cal D}\left[\varepsilon_k/M\right]  & 0 \\
\hline 0 &  {\cal D}\left[E_k\right]
& {\cal D}\left[RLC_k\right]  &  {\cal D}\left[M^{ps}_k\right] \\
\hline 0 & 0 &  {\cal D}\left[M^{ps}_k\right]  &  {\cal D}\left[{\cal L}^{si}_k\right]
\end{array}\right)\,,
\end{eqnarray}

\begin{eqnarray}
{\cal A}\equiv
\left(\begin{array}{c|c|c|c}{\cal I}_N  & - {\cal D}\left[\alpha_N\right]
\cdot {\cal B}_{Nk}\cdot {\cal D}\left[m_k/M\right]
& 0& 0 \\
\hline 0 & {\cal I}_k   & 0 & 0 \\
\hline 0 & 0 &{\cal I}_k  &  0 \\
\hline 0 & 0 & 0 &  {\cal I}_k\end{array}\right)\, ,
\end{eqnarray}

\begin{eqnarray}
{\cal Q} \equiv \{\xi_N, q_k, J_k, I_k-f^I_k\}\, ,
\quad{\cal F} \equiv \{f_N, f^t_k, f^p_k, f^s_k\}\, ,
\end{eqnarray}

where ${\cal I}_n$ is the $n$-dimensional identity matrix, the dynamical 
system of equations can be written in a concise way

\begin{eqnarray}\label{system}    
{\cal Z} \cdot {\cal Q}= {\cal A} \cdot {\cal F}\, ,
\end{eqnarray}

which in turn implies

\begin{eqnarray}
\label{findIk}    
\{\xi_N, q_k, J_k, I_k\}={\cal Z} ^{-1} \cdot  {\cal A} \cdot {\cal F}+
\{0, 0, 0, f^I_k\}\, .
\end{eqnarray}

This reduces the problem of finding the currents $I_k$, and their 
spectral correlation matrices, to the inversion of the matrix  ${\cal Z}$.

In some particular case, such inversion is very easy to compute.
For instance, for the case of 6 transducers in the TIGA configuration, if we 
include in the model only the first 5 quadrupolar sphere modes and the scalar 
one, then

\begin{eqnarray}
\label{Bort}
B_{Nk} \cdot B^T_{kN}=B^T_{kN} \cdot B_{Nk}=\frac{3}{2\pi}{\cal I}_6\, .
\end{eqnarray}

In this case, in the idealized hypothesis that all the 6 transducers and 
readout chains are identical, thus implying that all the blocks of the kind 
${\cal D}\left[X_k\right]$ appearing in ${\cal Z}$ are proportional to the 
identity, one obtains that the following redefinition

\begin{eqnarray}
{\cal Q}\equiv {\cal R} \cdot {\cal Q}'\,,
\end{eqnarray}

with

\begin{eqnarray}
{\cal R}
\equiv
\left(\begin{array}{c|c|c|c}{\cal I}_N  & 0  & 0 & 0 \\\hline 0 & B_{Nk}  & 
0 & 0 \\\hline 0 & 0 & B_{Nk}  &  0 \\\hline 0 & 0 & 0 &  B_{Nk}\end{array}\right)\, ,
\end{eqnarray}
leads to
\begin{eqnarray}
{\cal Z} \cdot {\cal Q}= {\cal R}\cdot {\cal Z}'\cdot {\cal Q}'\, ,
\end{eqnarray}
where all the 16, $6\times 6$ blocks of ${\cal Z}'$ are diagonal.

Since ${\cal R}$ is easily invertible by means of eq.(\ref{Bort}), the $I_k$'s 
are easily found. This particular case is nothing but a rephrasing of the idea 
of {\it mode channels} of an ideal TIGA configuration, see 
\cite{Merkowitz:1997qc}\footnote{The presence of the scalar mode is not 
necessary for the definition of mode channels, and indeed this is not present 
in the original derivation. We included it here because it makes the formalism 
more powerful, and for reasons that we will explain in the next subsection.}.

However, since our purpose is to simulate a realistic detector, we will not 
take advantage of any idealization, and we will invert the ${\cal Z}$ matrix 
numerically.

\begin{figure}
\begin{center}
\includegraphics[width=.65\linewidth]{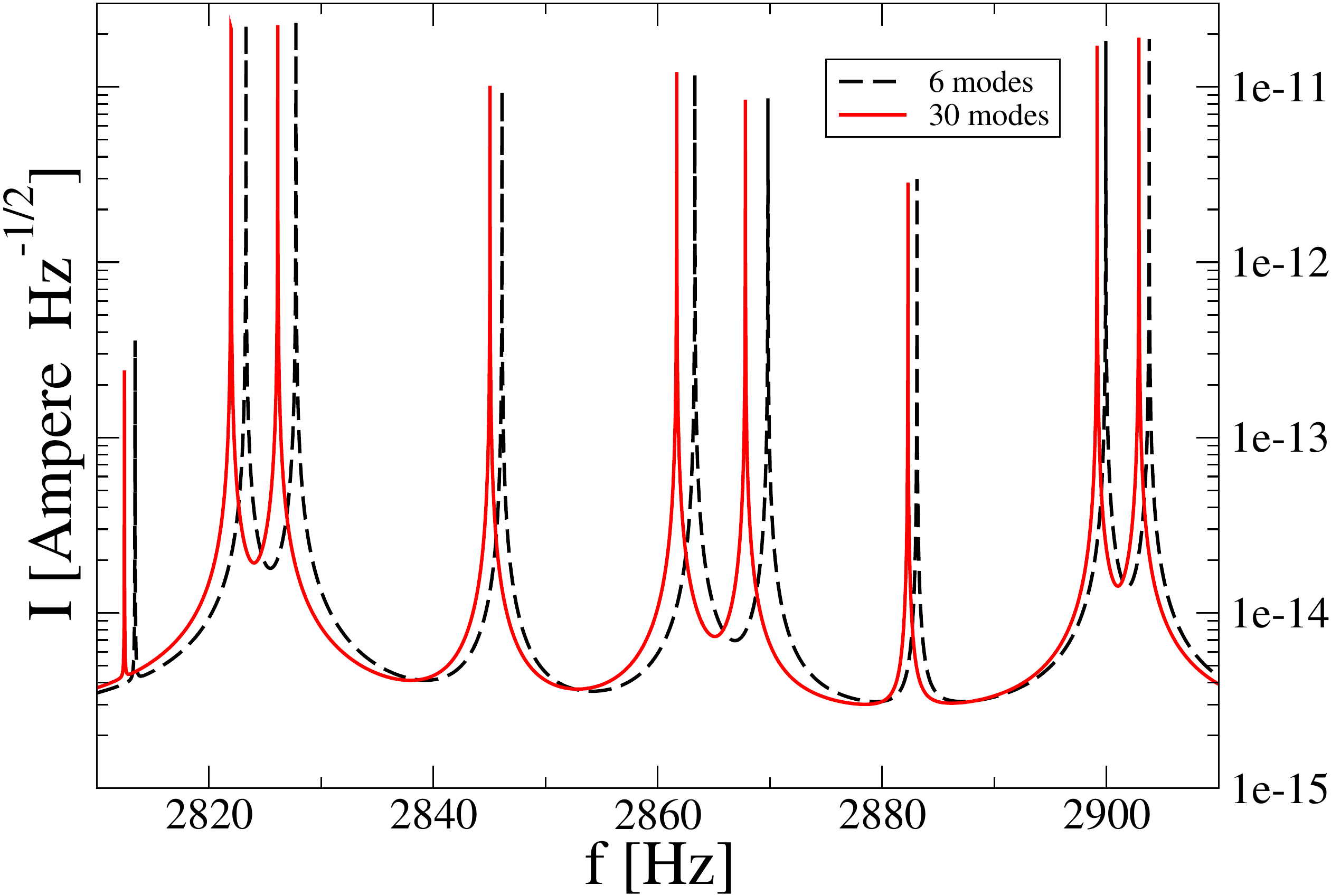}
\end{center}
\caption{Theoretical output of one of the transducers when 6 or 30 sphere
  vibrational modes are included in the model. 
To emphasize the differencies, a restricted region of the spectrum have been
displayed: this is the reason why  only nine out of the eleven resonant
modes are visible.}\label{6vs30}
\end{figure}

For the same reason we have decided to include many sphere modes in our model, 
contrarily to the common practice of considering only the lowest quadrupole 
multiplet, that is the one that interacts with the GW. More precisely, we have 
included in our model all the radial modes with resonant frequency below
that of the scalar mode: this gives a total of 30 degrees of freedom in this 
part of the system, distributed among one scalar, one vector, two quadrupolar, 
one {\cal l}=3 and one {\cal l}=4 multiplets.
The precise reason of this choice will be discussed in subsection 
\ref{subs_transfh}; for the moment we just mention that even if the extra
modes resonate far away the interesting region of the spectrum
(which is around $3$kHz), nevertheless their inclusion in the model affects
the response function even in the relevant spectral region, shifting the peaks 
in the response function of a few Hz's, as can be seen in fig.\ref{6vs30}.
In a real detector, however, such shift may be masked by the action of the
suspension system, finite-size effects of the transducers and/or some other 
unmodelled factors. At the moment we are not able to predict the 
magnitude of these unmodelled effects for miniGRAIL. 

For what concerns the transducers arrangement, we will stick to the TIGA 
configuration (with non identical transducers, however), although our analysis 
can easily be adapted to other configurations, provided the number of 
transducers is kept equal to six.

\subsubsection{From expectation values to a typical noise realization}
\begin{figure}
\begin{center}
\includegraphics[width=.65\linewidth]{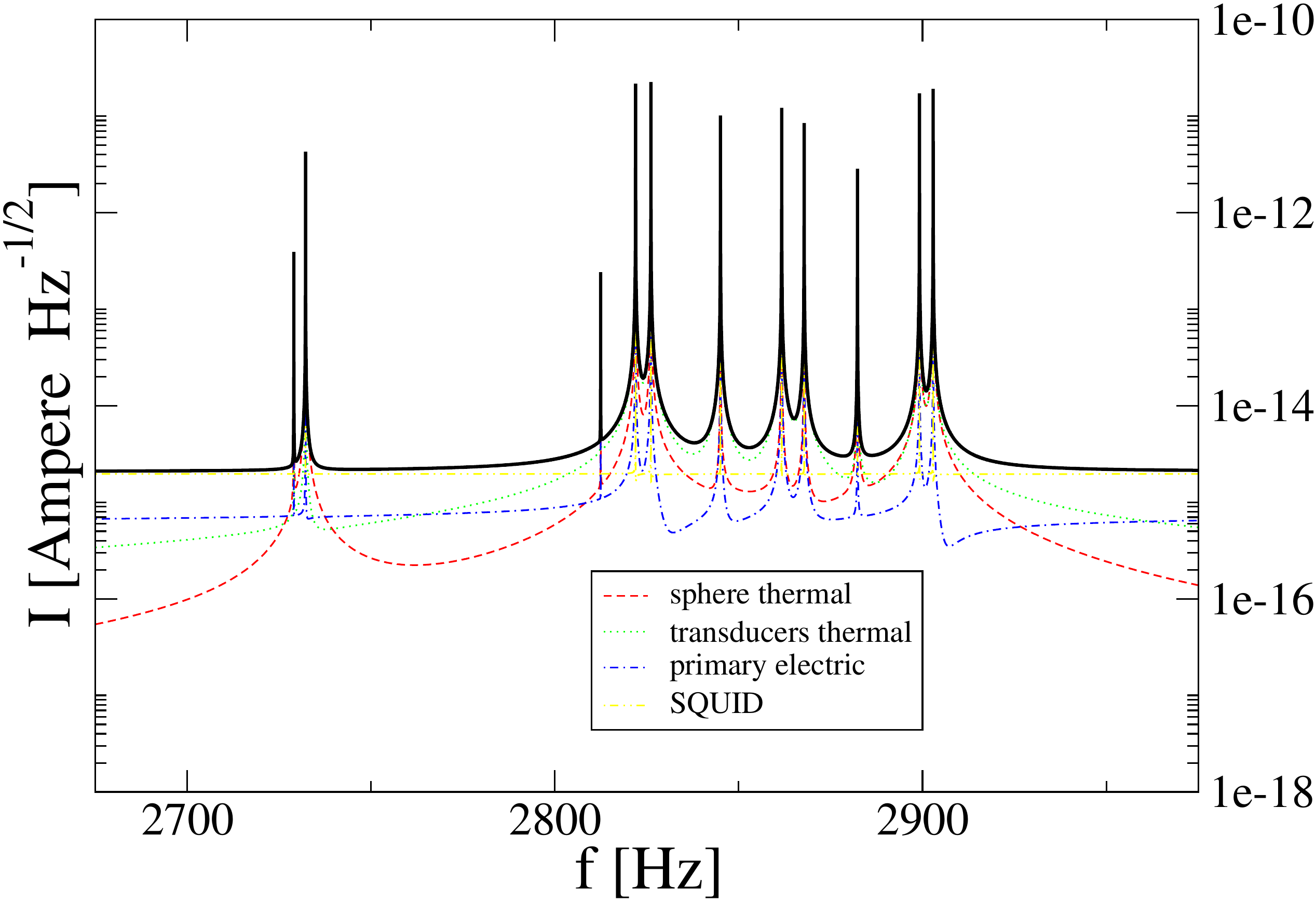}
\caption{The calculated output of the transducer \#0, along with the various
  noise contributions.}\label{tout0}
\vspace{1cm}
\includegraphics[width=.65\linewidth]{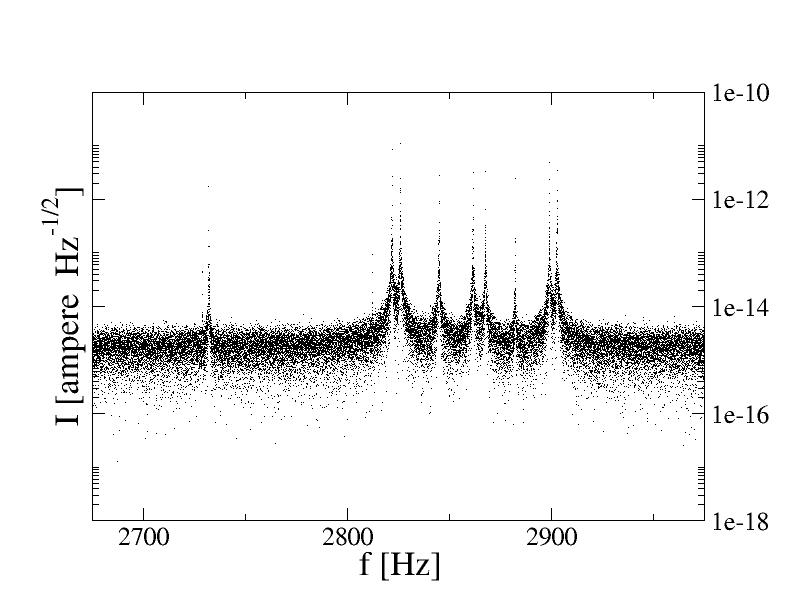}
\caption{Simulated output of transducer \#0, from \cite{Foffa:2008pm}.}\label{tout0_rec}
\end{center}
\end{figure}

~\\
Once ${\cal Z}$ has been numerically inverted, we can proceed to determine the 
currents $I_k$ through eq.(\ref{findIk}), and in particular we can 
estimate the corresponding spectral density matrix from 
eqns.(\ref{noisemt})-(\ref{noiseSQ}):
\begin{eqnarray}\label{noiseout}    
<I_k(\omega) I^*_{k'}(\omega')>\equiv 
S^I_{kk'}(\omega)\delta(\omega-\omega')\, .
\end{eqnarray}
The square root of one of the diagonal elements of $S^I_{kk'}$
is shown in fig.\ref{tout0}, along with the different noise contributions. 

This is still not enough for our purposes: what we really want to get at this 
stage is not just expectation values  and cross-correlations, but an 
{\it actual realization} of a possible detector output. To obtain this, we 
have generated several time series of white noise, we have taken their 
numerical Fourier transforms, and "colored" them according to the spectral 
dependencies of eqns.(\ref{noisemt})-(\ref{noiseSQ}), thus obtaining a 
possible realization of the Fourier transform of the various stochastic 
forces $f$'s.
Finally, we have derived, still in Fourier space, the realizations of
the currents $I_k$, by means of eq.(\ref{findIk}), satisfying 
eq.(\ref{noiseout});
one result of this procedure is displayed in fig.\ref{tout0_rec}.

We underline the fact that fig.\ref{tout0_rec} does not represent just a
white spectrum 
"colored" according to the expectation value shown in fig.\ref{tout0}, but it 
rather contains, in virtue of the procedure we have adopted to generate it, 
all the correct cross-correlations with the other output currents, i.e. the 
non-diagonal elements of eq.(\ref{noiseout}). This fact is crucial if 
the injected signals are to be reconstructed.

To summarize the results of this subsection, we have numerically generated
the six current outputs of a realistic sphere working 
in the TIGA configuration. Our model is realistic in the sense that we have 
chosen the values of all the components of the simulated detector according to 
what is actually being implemented on miniGRAIL, and we have included an 
adequate number of sphere vibrational modes.

\subsection{Interaction with the GW: the transfer matrix}\label{subs_transfh}
We now change point of view: we assume that the six secondary currents
are given and we would like to determine the sensitivity of our detector
to GW's.
In other words, we are trying to reproduce as faithfully as possible a real 
experimental situation.

We remind that when a GW impinges the detector, a deterministic force is added 
to the quadrupolar components of $f_N$:

\begin{eqnarray}\label{redfN}
f_N\rightarrow f_N+\frac{1}{2}R\chi\ddot{h}_N\quad{\rm for}\quad N=
0,\dots,4\, ,
\end{eqnarray}
where $R$ is the sphere radius, $\chi=-0.3278$ for CuAl6\%
(the alloy miniGRAIL is made of) and the $h_N$'s are the quadrupolar spheroidal
components of the GW, defined as 
follows through decomposition via the tensor spherical harmonics:

\begin{eqnarray}\label{hN}
h_{ij}=\sum_{N=0,4}{\cal Y}^{(2)}_{m_N, ij} h_N\, ,
\end{eqnarray}
with $m_N=0,1c,1s,2c,2s$ and
$Y^{(2)}_{m_N}=\sum_{i,j}{\cal Y}^{(2)}_{m_N, ij}n^i n^j$, being $n^i$ the 
versor of the arrival direction of the wave.

\subsubsection{Issues in getting the $h_N$'s from the transducer outputs}
~\\
The problem is clearly to get the most possible information about the $h_N$'s 
from the knowledge of the six $I_k$'s. For this goal, it would be enough to 
determine, and most of all to invert somehow, the ${\cal T}_{kN}$ matrix, 
which is defined as the $k\times 5$ block determined by crossing the last $k$ 
lines with the first $5$ columns of the matrix ${\cal Z}^{-1}{\cal A}$ which 
appears in eq.(\ref{findIk}).
Basically, ${\cal T}_{kN}$ is the matrix entering the relation:
\begin{eqnarray}\label{sysIfN}
I_k={\cal T}_{kN}\cdot f_N\,.
\end{eqnarray}
with $f_N$ given by eq.(\ref{redfN}), which holds in the limit of very large 
SNR.

In general, the solution to this problem is very much dependent on the number 
of transducers. If such number is less than four, then there are not enough 
experimental inputs to determine all the $h_N$'s, and thus the matrix $h_{ij}$ 
cannot be completely reconstructed. Rather, only some linear
combinations of the $h_N$'s (or, in another notation, of the GW polarization 
components) can be determined\footnote{This situation is pretty similar to what
happens with interferometers and resonant bars, where only one output is 
present.}.

If there are four transducers, then in principle the metric perturbation can be
reconstructed, provided one imposes a priori on the data the constraint that 
the five $h_N$'s 
should form a transverse tensor. The five transducers case is similar, with 
the advantage that one can use the transversality condition as a veto, thus 
reducing the false alarm rate at a given SNR, rather then imposing it from the 
beginning.

If the transducers are six, as in the present case, the situation is more 
involved because the system is over-constrained by the fact that the $5$ 
$h_M$'s are to be determined by the $6$ outputs.
This kind of problem has generally no solution, as it is physically correct 
since the $I_k$'s are not determined {\it only} by the GW, but also by the 
detector noise; however in some cases an optimal solution can be found.

First of all, we have seen that in an ideal TIGA configuration (identical 
readout chains) a change of basis brings the matrix ${\cal Z}'$ to a block 
diagonal form. By inspection of the form of the system
of equations (\ref{system}) it can be seen that this implies also that
the ${\cal T}_{kN}$ matrix has the following pseudo-diagonal form:
\begin{eqnarray}\label{easyT}
{\cal T}_{kN}\equiv
\left(\begin{array}{ccccc}* & 0 & 0 & 0 & 0 \\0 & * & 0 & 0 & 0 \\0 & 0 & * & 0 & 0 \\0 & 0 & 0 & * & 0\\
0 & 0 & 0 & 0 & * \\0 & 0 & 0 & 0 & 0\end{array}\right)\, ,
\end{eqnarray}
the last row of zeroes meaning that the sixth redefined component of $I'_k$ 
{\it is independent} of the
$h_N$'s. In this case one has just to consider the square, and invertible, 
matrix made of the first $5$
lines of ${\cal T}_{kN}$: i.e. only the first five $I'_k$'s (which are 
nothing else but the TIGA mode
channels) are used to find the five $h_N$'s and the system is again well 
determined.

When the idealized conditions for a perfect TIGA configuration no longer hold,
the situation becomes more
complicated because there is no simple way to get to the form (\ref{easyT}) 
for the ${\cal T}_{kN}$ matrix; in particular, in the general case all the six 
current outputs (or any linear combination of them) depend on the GW 
parameters.

In this case, the system (\ref{sysIfN}) is really over-constrained and has in 
general no exact solution; in such cases, the best one can do from the 
mathematical point of view is to find the best possible approximate solution 
for the system, that is, in the case at hand, a vector $f^*_N$ such that
the norm of the vector
\begin{eqnarray}\label{minnorm}
I_k-{\cal T}_{kN}\cdot f^*_N\, ,
\end{eqnarray}
is minimal.
If the matrix ${\cal T}_{kN}$ has maximum rank, then the pseudo solution
$f^*_N$ can be shown to be unique, and to be given by
\begin{eqnarray}\label{pseudoinv}
f^*_N=({\cal T}^T_{Nk}\cdot {\cal T}_{kN})^{-1}\cdot {\cal T}^T_{Nk}
\cdot I_k\, .
\end{eqnarray}
This strategy, which has been proposed for example in \cite{Merkowitz:1997qs} 
is certainly a good one in the case of very high SNR, while at low SNR, and in 
particular in the determination of the detector's sensitivity (where $SNR=1$ by
definition), it shows some drawbacks.

The problem is that, when the noise gives a non negligible contribution to the 
output currents, there is no physical reason to expect that the actual $h_N$'s
are the ones that minimize the norm of the vector (\ref{minnorm}).

\subsubsection{Inclusion of the scalar mode}
~\\
Our proposal in this case is to include the scalar vibrational mode of the 
sphere, so that eqns.(\ref{redfN}), (\ref{hN}) are now turned into
\begin{eqnarray}\label{newredfN}
f_N\rightarrow f_N+\frac{1}{2}R\chi_N\omega^2 h_N\quad{\rm for}\quad 
N=0,\dots,5\, ,
\end{eqnarray}
\begin{eqnarray}\label{newhN}
h_{ij}=\sum_{N=0,4}{\cal Y}^{(2)}_{m_N, ij} h_N + \frac{{\cal I}_{ij}}
{\sqrt{4 \pi}} h_5\, ,
\end{eqnarray}
being ${\cal I}_{ij}$ the identity matrix, and $\chi_N=\chi$
for $N=0,\dots,4$, while it can be shown that $\chi_5=-3.8$.

As we have already discussed, inclusion of the scalar mode makes the position 
matrix $B_{Nk}$ invertible and orthogonal, see eq.(\ref{Bort}), because the 
row corresponding to the scalar mode is orthogonal to the $5$
lines related to the quadrupolar modes. This means that adding the scalar 
mode is the best possible way to complete the $5$ quadrupole position vectors 
to a basis of the six-dimensional linear space.

If we proceed this way, another sixth column must be included in the 
definition of ${\cal T}_{Nk}$, which now becomes an invertible $6\times 6$ 
matrix. In other words, including the scalar mode is a way to turn an 
over-constrained rectangular system into a squared one; it is clear that the 
same result could have been obtained by adding any other sphere vibrational 
mode rather than the scalar, but as previously stated, inclusion of the scalar 
mode makes the system particularly symmetric.

A word of caution is needed here: by the inclusion of the scalar mode, and in 
particular its parameterization by eq.~(\ref{newhN}),
we do not aim at detecting scalar gravitational radiation.
We are rather parametrising the noise contained in the scalar vibrational
mode of the sphere, and dropping it from the analysis of the quadrupolar
modes: in other words it makes much sense to take the part 
of noise related to the trace out of the 5 quadrupolar $h_N$'s, and the 
addition of the scalar mode, which is sensitive exactly to this part, is the 
way to do it.

Stated in yet another way: the system (\ref{sysIfN}) is over-constrained, 
and one needs an extra input to obtain the best possible solution for $h_N$.
In the idealized TIGA case, this extra input comes from the existence of 
mode channels, while in the high SNR case a good criterion is the minimal 
distance one. 
In general, we think that a physically motivated option is to express the 
six noisy output currents in terms of the five quadrupolar sphere modes plus 
the scalar one, that is the best way to complete the quadrupolar modes to a 
basis in the six-dimensional space, and to ensure that the trace part of the 
noise does not ``leak'' into the quadrupole modes.

\subsubsection{Transfer matrix}
~\\
We now come to the determination of the transfer function and of the GW
components. The system (\ref{sysIGW}) can be rewritten as follows
\begin{eqnarray}\label{sysIGW}
I_k(\omega)=\frac{1}{2}R\omega^2 {\cal T}_{kN}(\omega)\cdot{\cal D}
\left[\chi_N\right]\cdot h_N(\omega)\, ,
\quad N,k=0,\dots,5\, ,
\end{eqnarray}
and numerically inverted (for every value of $\omega$) to give 
\begin{eqnarray}\label{transf}
h_N(\omega)= T_{Nk}(\omega)\cdot I_k(\omega)\, ,
\end{eqnarray}
where $T_{Nk}(\omega)$ defines the  {\it transfer function} (which in this 
case is actually a transfer matrix) of the system.

\begin{figure}
\begin{center}
\includegraphics[width=.45\linewidth]{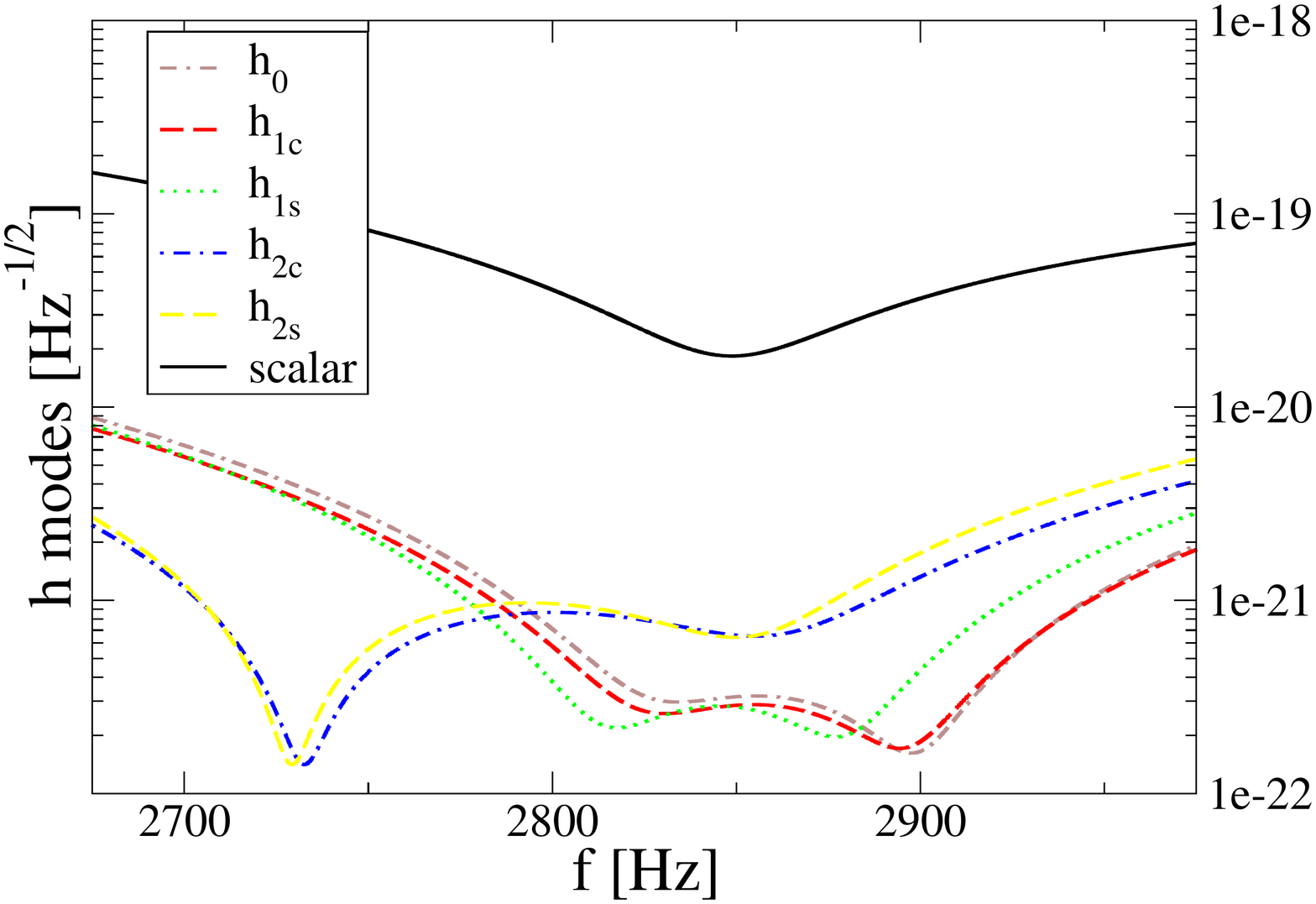}
\hspace{.3cm}
\includegraphics[width=.45\linewidth]{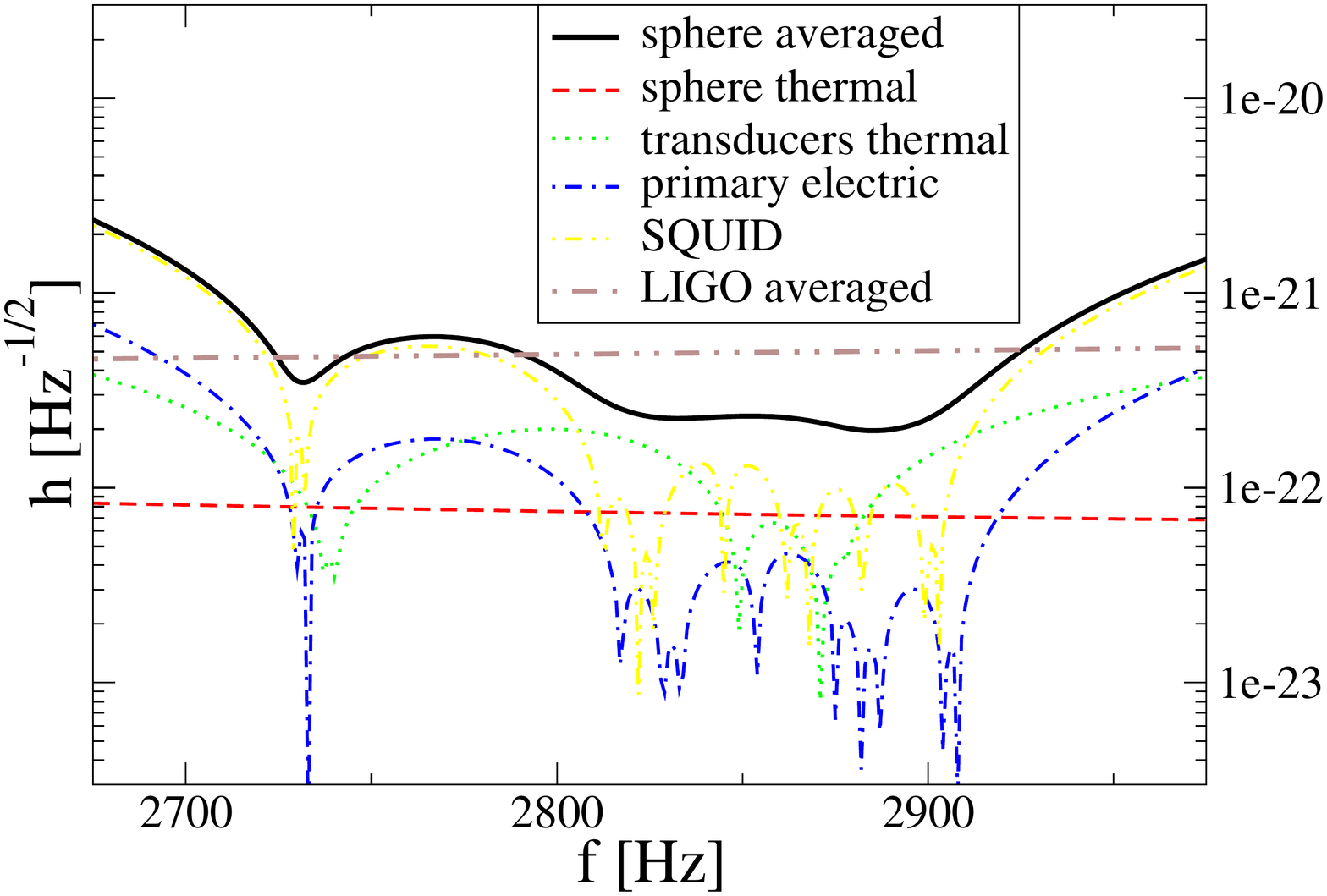}
\caption{On the left, all the GW quadrupolar modes and the scalar one.
On the right, the sky-averaged sensitivity of the sphere
(along with the various noise contributions), compared to that of LIGO.}
\label{hout}
\end{center}
\end{figure}

Eq.(\ref{transf}) can be combined with the results of the previous section, 
in order to give the $h_N$ spectral density matrix, which can be used to 
estimate the detector sensitivity.
The left part of fig.~\ref{hout} shows the diagonal components of such
sensitivity matrix, that is square root of the expectation values
$<h_N h^*_N>$, for all values of $N$.
Since the different modes have different sensitivity curves, and since they
are also cross-correlated, the overall sensitivity of the sphere is not
isotropic. On the right part of the same figure we have shown such sensitivity
averaged over all the incoming directions and over all the possible
linear polarizations, together with the various noise contributions,
and we have compared it with the corresponding quantity for the LIGO
interferometer \footnote{Strictly speaking, the sky average of an
interferometer strain sensitivity is infinite because of the presence of
a blind direction. Here we have divided the
sensitivity for optimal direction and polarization by the square root
of the angular efficiency factor $F=2/5$ for an interferometer.}.

It can be noticed that i) the five quadrupolar components
having different sensitivities is due to the fact that the readout chains
are not identical and that the mode resonances are not degenerate, and ii)
as expected the detector is poorly sensitive to the scalar mode: this happens 
because the transducers are tuned to the quadrupolar multiplet 
whereas the scalar mode resonance is at a much higher frequency 
then the quadrupole (roughly at $5$kHz).
 
Coming from expectation values to our detector simulation, the same procedure 
can be applied to the Fourier transform of the $I_k$'s realization produced in 
the previous subsection, to give the Fourier transform of
the $h$ modes.

\begin{figure}
\begin{center}
\includegraphics[width=.65\linewidth]{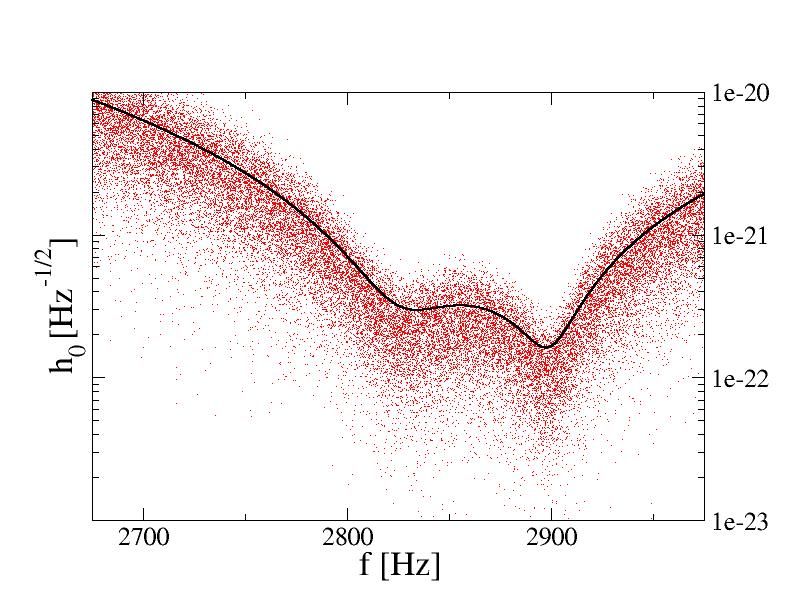}
\caption{The mode $h_0$ as it has been reconstructed starting from an actual
noise realization, superimposed with its expectation value.}\label{hout0_rec}
\end{center}
\end{figure}

For example, the quadrupolar $h_0$ mode spectrum is shown in
fig.~\ref{hout0_rec}, together with its expectation value.

Once the $h_N(\omega)$'s are transformed back in the time domain, six time 
series are obtained, which describe the noise corresponding to the  
components of the $5$ spheroidal modes of the tensor $h_{ij}$ and of the scalar
mode, see eq.~(\ref{newhN}). However, given the poor sensitivity of the
detector to the latter (see fig.~\ref{hout}), this channel is bound to contain
instrumental noise only and it will be dropped in the forthcoming analysis. 

\vskip 2pt

\section{The analysis pipeline}
\label{pipeline}
We have now the five time series corresponding to the quadrupolar modes,
which are enough and necessary to reconstruct the most general 
\emph{symmetric, traceless} $3\times3$ matrix.

Still this is a redundant description of a GW, and we will exploit this 
redundancy to discriminate between real GW signals and excitations of the
modes due to disturbances other than gravitational.

The treatment in this section closely follows the exposition of 
\cite{Foffa:2008pm}, which is summarized here.

\subsection{The scalar trigger}
If one knew exactly the form of the expected signal, the optimal strategy would
be to perform a multidimensional matched filter as follows:
\be
h_N(\omega)\rightarrow (h^{sig}\cdot h)_S\, ,
\ee
where the $h^{sig}_{N}(\omega)$'s are the Fourier modes of the quadrupolar
components of the expected signal, and we have introduced the following
notation:
\be
(X\cdot Y)_S\equiv X^{\dag}_N(\omega)\cdot(S^h)^{-1}_{NN'}(\omega)\cdot 
Y_{N'}(\omega)\quad N,N'=0\, ,\dots\, ,4\, .
\ee
However in a realistic case neither the temporal dependence of the signal,
nor the arrival direction are known. One could think of overcoming this
difficulty by doing several matched filters with signal of various shapes
and polarizations, and coming from several directions in the sky, but this
method is computationally very intensive, and in addition results in a
much higher false alarm rate than the method we are going to describe now.

We would like to find triggers out of the stretch of data, without need of a
detailed knowledge of the signal.
This has been done in \cite{Foffa:2008pm} by building the following quantity:
\be \label{Hdef}
H(\omega) \equiv (h\cdot h)_S\,.
\ee
The Fourier transform of this quantity is then fed to a suitably
adapted version of WaveBurst,
the event trigger generator used in the LIGO data analysis \cite{Klim,Klim2}.
The WaveBurst algorithm make use of the wavelet decomposition, and among the 
bank of wavelet packets we picked the Symlet base with filter length sixty
\cite{Mal}.

If a GW signal is present in the $h_N$'s, WaveBurst will detect it in $H$
whatever its shape, polarization and arrival direction, provided of course,
that it is strong enough.

Once the trigger has been established the analysis is performed on the modes 
$h_N$'s, by collecting for each trigger the values of the wavelet
coefficients. At this point the arrival direction can be reconstructed by a 
likelihood method analysis. This is different than what is being
done by the Coherent WaveBurst algorithm \cite{Klimenko:2008fu}, where the 
likelihood value is
computed at every point of the time-frequency plane, as we do not have the
sufficient computational power to perform such a daunting task.
On the other hand, as it will be shown in sec.~\ref{result}, adopting the 
above mentioned scalar trigger allows a good detection efficiency and reduce 
enormously the computational cost of the procedure, as our algorithm runs on
a standard MacPro machine with two 3GHz processors and it takes for the 
analysis a time which is roughly two thirds of the actual time duration of the 
data. 

The values of $\theta,\phi$ of the GW arrival direction are then found as the 
ones maximizing the likelihood function.
This method identifies an arrival direction of the candidate GW event no
matter if a real GW has excited the detector or a glitch, say, has taken place.
To confirm this direction determination we combine it with a different, 
algebraic method. When the two methods do not determine the same directions, 
within some tolerance to be discussed quantitatively in sec.~\ref{result}, the 
event can be discarded as spurious. 

\begin{figure}
\begin{center}
\includegraphics[width=.3\linewidth,angle=90]{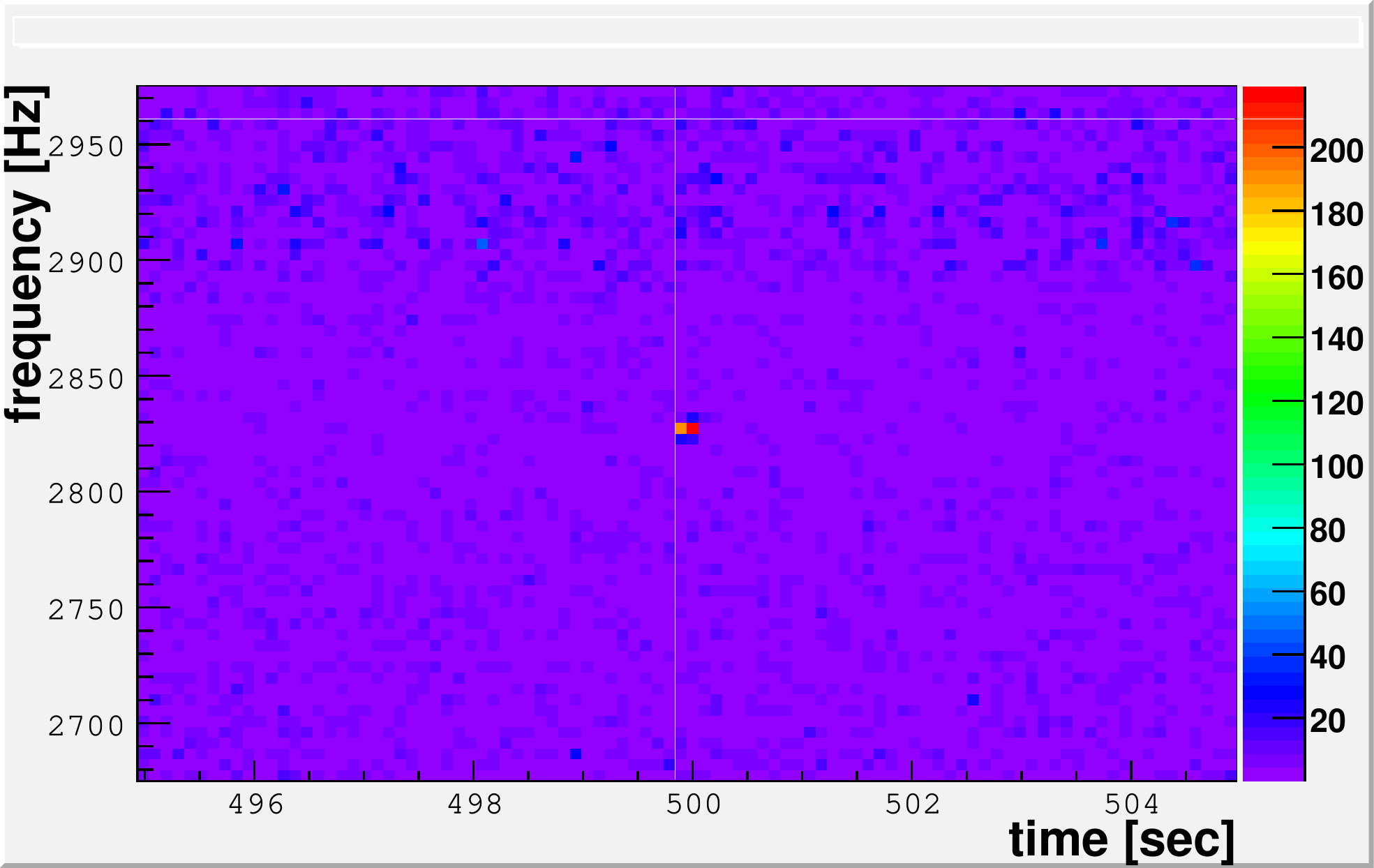}
\includegraphics[width=.3\linewidth,angle=90]{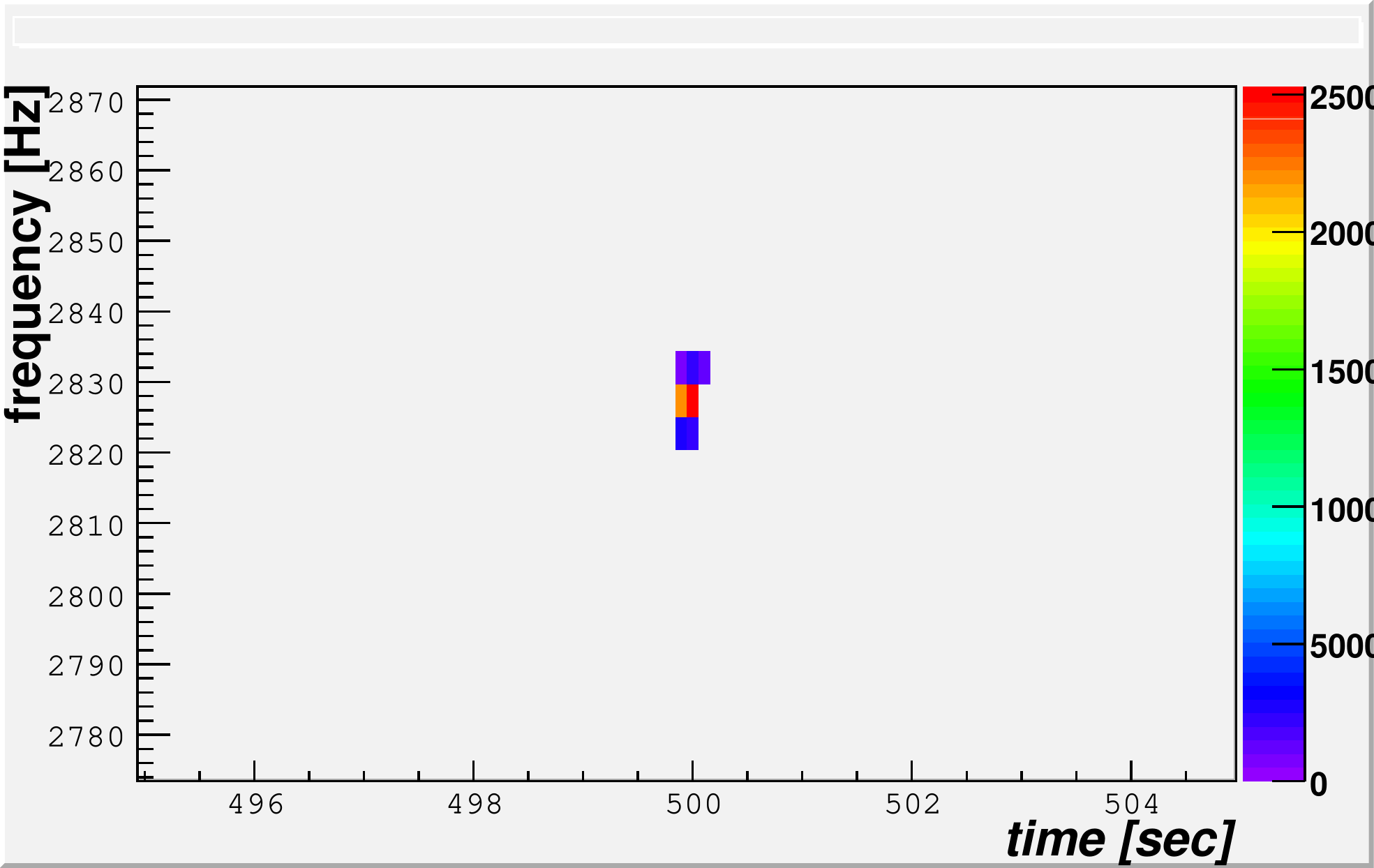}
\end{center}
\caption{Example of event trigger. Sine-gaussian GW injection, centered at 
$t=500$sec, width $50$msec, central frequency $2825$Hz and strength 
$h_{rss}=7\times 10^{-21}{\rm Hz}^{-1/2}$. On the left is the $H$ mode defined 
in eq.~(\ref{Hdef}), and on the right the likelihood of eq.~(\ref{elle_acca}) 
on the points of the trigger.} 
\label{ex_trigger}
\end{figure}

\subsection{Likelihood method for direction reconstruction}

The probability of having a given stretch of data $\{h_N\}$ assuming a GW with 
polarization amplitudes $h_+$ and $h_\times$ from the direction identified
by the usual polar angles $\theta,\phi$, is often referred as likelihood
function and is denoted by $p(\{h_N\}|H_{\zeta_N})$.
For stationary, Gaussian, white noise with zero mean, and by taking into
account that the noises in the different channels are correlated,
such probability is
\be
p(\{h_N\}|H_{\zeta_N})\propto\exp\paq{-((h-\zeta)\cdot(h- \zeta))_S/2}\,,
\ee
where $\zeta_N$ is defined as
\be
\zeta_N=F_N^{+}(\theta,\phi)h_{+}+ F_N^{\times}(\theta,\phi)h_{\times}\,,
\ee
$F_N^{+,\times}$ being the pattern function of the $N$ mode for the 
$+,\times$ polarization.

Following the standard procedure \cite{Flanagan:1997kp,Anderson:2000yy}, 
the likelihood ratio $\Lambda$ can be defined as
\be
\label{loglike}
\Lambda(h_{+,\times},\theta,\phi)=
\frac{p(\{h_N\}|H_{\zeta_N})}{p(\{h_N\}|H_0)}\,.
\ee
By maximizing the likelihood ratio eq.~(\ref{loglike}) with respect to 
$h_+,h_\times$ a function $\mathcal L(\theta,\phi)$ of the angles alone is 
found
\renewcommand{\arraystretch}{1.8}
\be \label{elle_acca}
\ba{l}
\ds
\!\!\!\!\!\!\!\!\!\!\!\!\!\!\!\!\!\!\!\!
\mathcal L_h(\theta,\phi)= \frac 12
\paq{(F^+\cdot F^+)_S(F^{\times}\cdot F^{\times})_S-(F^+\cdot
F^{\times})^2_S}^{-1}\times\\
\!\!\!\!\!\!\!\!\!\!\!\!\!\!\!\!\!\!\!\!
\ds \paq{F^{\times}\cdot F^{\times})_S (h\cdot F^+)^2_S+
(F^+\cdot F ^+)_S(h\cdot F^{\times})^2_S-
2(h\cdot F^+)_S(h\cdot F^{\times})_S(F^+\cdot F^{\times})_S}\, .
\ea
\ee
\renewcommand{\arraystretch}{1}
Maximizing the likelihood ratio is equivalent to maximizing the 
\emph{posteriori} probability $p(H_{\zeta_N}|\{h_N\})$ in the case of flat 
priors. We then take the values of $\theta,\phi$ (which enter the expression 
for the $F_N$'s) maximizing $\sum_{\{trigger\}} \mathcal L_h(\theta,\phi)$,
that is the sum of the likelihood over every point of time-frequency
plane exceeding the threshold, as our estimation of the arrival direction of 
the candidate event.\footnote{Our (\ref{elle_acca}) is equivalent to (33) of 
\cite{Klimenko:2005xv}. There it is further introduced a parameter $\epsilon$
which is crucial to define an \emph{improved} likelihood in the case 
$\epsilon\ll 1$. In our case $\epsilon \lesssim 1$.}

\subsection{Determinant method for direction reconstruction}
\label{subsdete}

Another method to reconstruct the direction of arrival GW by exploiting 
the (redundancy of the) five quadrupolar modes is based on linear algebra 
considerations \cite{Merkowitz:1997qc}.
A general metric perturbation in the transverce-traceless gauge
is necessarily parametrized by a symmetric, traceless $3\times 3$ matrix with 
zero determinant. No matter which arrival direction nor polarization,
the \emph{null eigenvector} always points to the arrival direction (with the
usual ambiguity in up and down direction).

As the quadrupolar modes allow full reconstruction of the metric perturbation,
we have an independent method to identify the incoming direction of the
signal.\\
Of course none of the eigenvalues is expected to be exactly zero at every point
in the trigger, we then proceed to order the three eigenvalues $\lambda_i$ so 
that $|\lambda_0|\leq |\lambda_1|\leq |\lambda_2|$ and define the following
quantity
\be \label{rdef}
r\equiv\frac{\sqrt 2|\lambda_0|}{\sqrt{\lambda_1^2+\lambda_2^2}}\,.
\ee
For a perfect GW-like $r$ vanishes, thus the smallest it is, 
the less the noise is contaminating the GW signal.
If for at least one point in the trigger $r<r_0$, with $r_0$ a suitably chosen
threshold, a direction can be identified by associating to the signal the 
direction designated by the eigenvector relative to the smallest eigenvalue 
$\lambda_0$.

For each point of the trigger satisfying the condition $r<r_0$ a 
direction is identified; the average is then taken by weighting them
with the factor $1/r$ in order to obtain a single direction for each trigger.

\section{Test of the method}\label{result}
\subsection{Setup and calibration}
In order to test the efficiency of our method, we injected
mock GW signals in the five channels corresponding to
the spheroidal components of $h_{ij}$ (see subsection \ref{subs_transfh}).

As efficiency is a good measure of the validity of the method only in presence
of an estimate of the false alarm rate, we also injected non-GW
signals which are supposed to mimic the presence of non-Gaussian
noise in the data.

In both cases the amplitude shape is a sine Gaussian
\be \label{sgaus}
h_{inj}(t+t_0)=h_0e^{-t^2/(2\Delta^2)}\sin(2\pi ft)
\ee
where $\Delta=0.05$sec, $f=2825$Hz, $t_0$ is the signal center and $h_0$ is 
determined by the signal $h_{rss}$.

\begin{figure}
\begin{center}
\includegraphics[width=.65\linewidth]{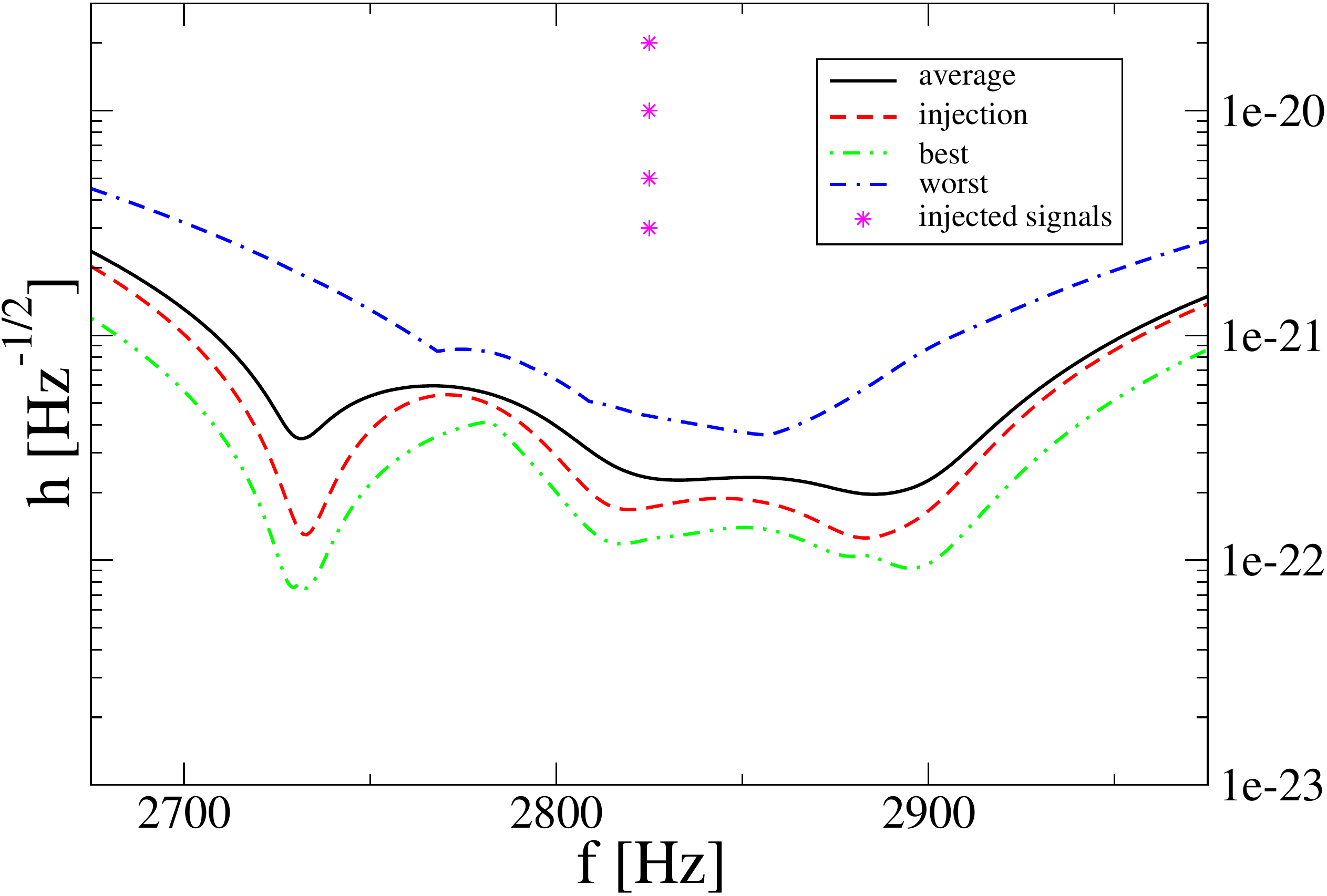}
\end{center}
\caption{The stars indicate the injections, while the dashed line is the strain
sensitivity for the arrival direction of the injections.
For comparison, we have displayed also also the averaged sensitivity
(continuous line), as well as the ones corresponding, at any given
value of the frequency, to the best and worst possible directions.}
\label{hout_inj}
\end{figure}

Four sets of injections (of 50 signals each) have been done
both for GW and non-GW signals, with amplitudes
$h_{rss}=\{3, 5, 7,10\}\cdot
10^{-21}{\rm Hz^{-1/2}}$, corresponding to $SNR\simeq 16,27,37,53$ in
amplitude, see figure \ref{hout_inj}.

For the mock GW signals, the amplitude distribution among the different
channels corresponds to linearly polarized GW's (either $+$ or $\times$)
with arrival direction $\theta=54.736^{\circ},\phi=45^{\circ}$, while
for non-GW signals the amplitude is distributed randomly on the different
channels and normalized to ensure the required $h_{rss}$.

For each injection detected by the trigger, the arrival direction has been
reconstructed by the two methods (likelihood and determinant)
explained in sec.~\ref{pipeline} and the event have been selected only when
the two methods had a discrepancy below some threshold $\delta$.

\begin{figure}
\begin{center}
\includegraphics[width=.75\linewidth]{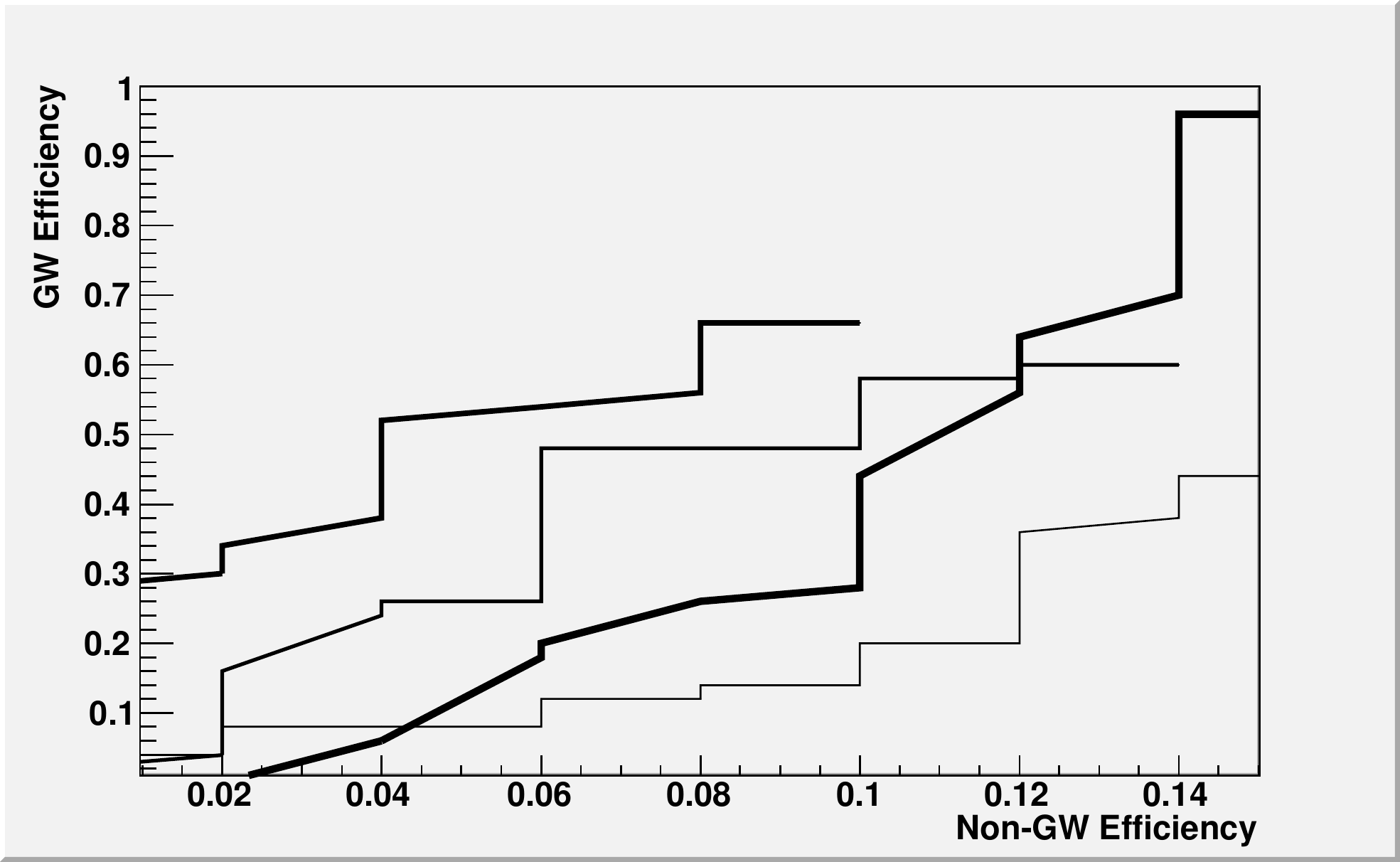}
\caption{Detection efficiency of GW-injections vs. detection efficiency of 
non-GW signals obtained by varying the maximum allowed distance between 
the two different direction reconstruction methods. Lines of increasing
thickness correspond to injections sets of increasing amplitude, 
$h_{rss}=\{3,5,7,10\}\times 10^{-21}{\rm Hz}^{-1/2}$.}
\label{roc_d}
\includegraphics[width=.75\linewidth]{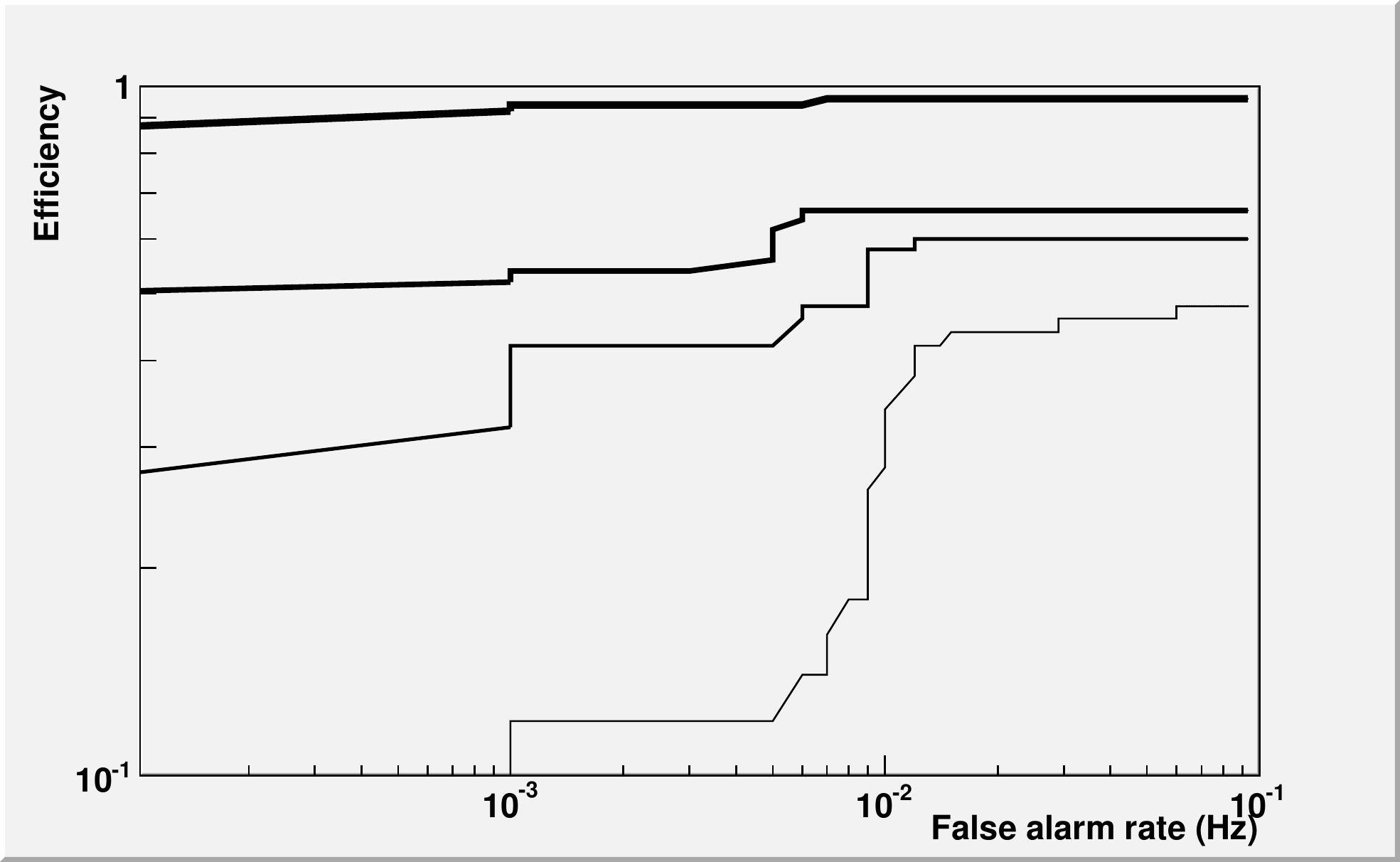}
\caption{Detection efficiency of GW-injections vs. false alarm rate of events 
on the data with just intrinsic detector noise, obtained by varying the
maximum allowed distance between the two different direction reconstruction 
method. Lines of increasing thickness correspond to injections sets of 
increasing amplitude, $h_{rss}=\{3,5,7,10\}\times 10^{-21}{\rm Hz}^{-1/2}$.}
\label{roc_dw}
\end{center}
\end{figure}

A ROC curve can be constructed to assess what is the best value for $\delta$:
for each set of injections one can define the efficiency as the fractional
number of injections which are recovered and for which the two 
different direction identification methods give directions separated by less 
than $\delta$.
By measuring the efficiency for GW and non-GW injections
and plotting the former vs the latter for different values of $\delta$,
fig.~\ref{roc_d} has been obtained, where the four curves refers to the four
different signal strengths.

The same can be done comparing the GW detection efficiency
with the false alarm rate obtained analyzing the data containing only
Gaussian noise (i.e. without any injection): the plot is reported in
fig.~\ref{roc_dw} and shows that, contrarily to what happens in
fig.~\ref{roc_d}, the ROC curve shows a strong dependence on $\delta$,
especially at low $h_{rss}$ (notice that the scales of the two figures
are different).

This calibration enabled us to set at $\delta_{thr}=0.2$ rad as the optimal
value; all the results reported in the remaining part of the paper have
been obtained with $\delta=\delta_{thr}$.

\subsection{Results} 

\subsubsection{Likelihood only} 
\begin{figure}
\centering
\includegraphics[width=.65\linewidth,angle=90]{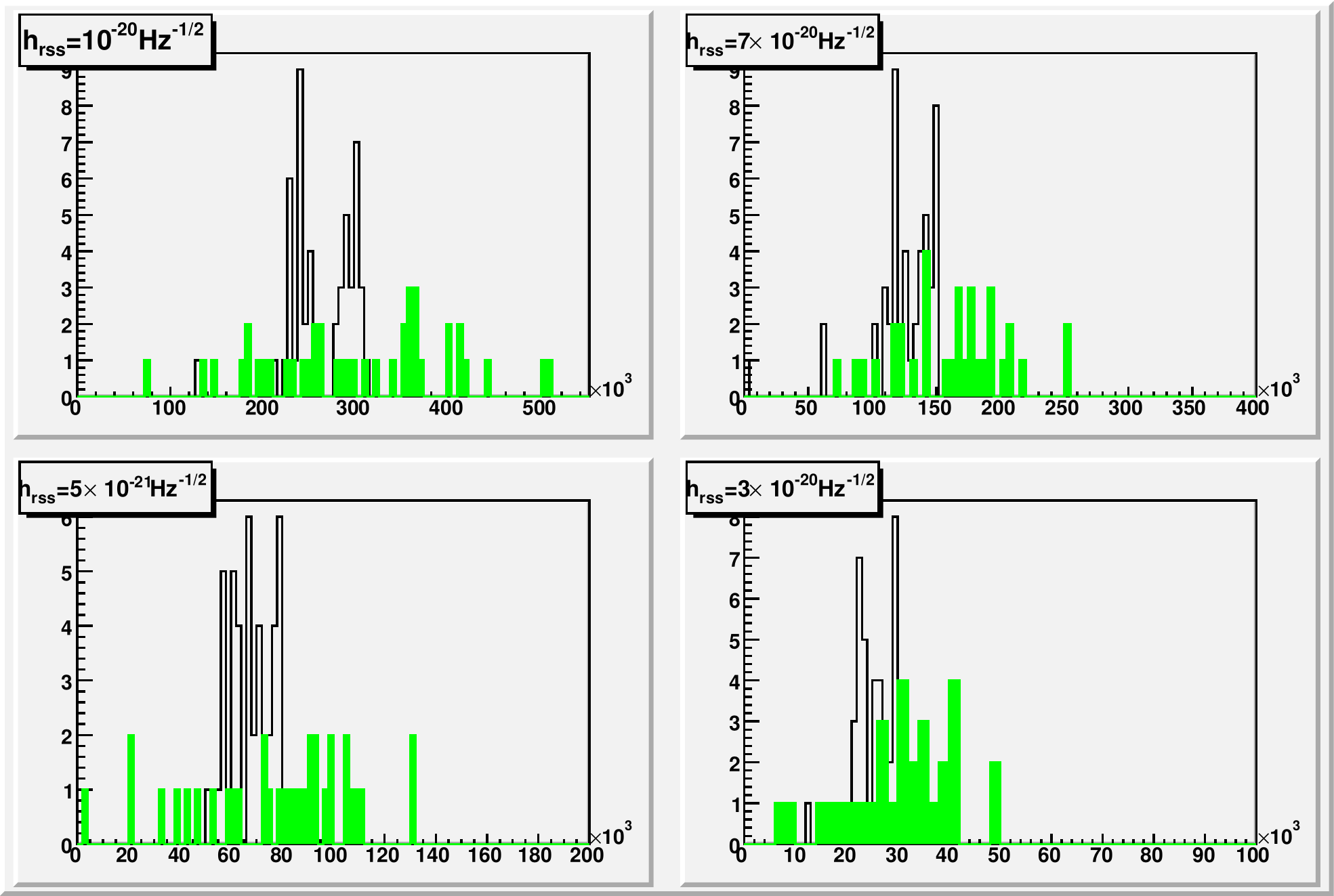}
\caption{Distribution of the values of the likelihood function for 
GW-injections (empty-black) and for non-GW injections (filled-green)
for various $h_{rss}$.}
\label{like_4}
\end{figure}

~\\
As a first step, we want to check what is the significance of finding
a trigger on the scalar mode $H$ defined in eq.~\ref{Hdef}.
In Fig.~\ref{like_4} we report the distribution of the likelihood values for
the triggers obtained in correspondence of GW and non-GW injections. 
Mean values and sigmas for the likelihood are reported in tab.~\ref{thlike}.

We see that our algorithm detects more efficiently GW injections
rather than non-GW ones, but at this stage it is impossible to discriminate
a priori between the two kinds of injections.

\begin{table}
\begin{center}
\begin{tabular}{|c|c|c|c|c|c|c|}
\hline
$h_{rss} {\rm (Hz^{-1/2})}$ & \multicolumn{3}{|c|}{GW} & 
\multicolumn{3}{|c|}{non-GW} \\
\cline{2-7}
& \#$_{inj}$/\#$_{trigg}$ & $\mu$ & $\sigma$ &\#$_{inj}$/\#$_{trigg}$ & $\mu$ & $\sigma$ \\
\hline
$10^{-20}$& 50/50 & $2.6\times 10^5$ & $3.5\times 10^4$ & 41/50 & $3.0\times 10^5$ & $9.8\times 10^4$\\
\hline
$7\times 10^{-21}$& 48/50 & $1.3\times 10^5$ & $2.7\times 10^4$ & 33/50 & $1.6\times
10^5$ & $4.3\times 10^4$\\
\hline
$5\times 10^{-21}$& 45/50 & $6.7\times 10^4$ & $9.5\times 10^3$ & 34/50 & $7.7\times 10^4$ & $3.1\times 10^4$\\
\hline
$3\times 10^{-21}$ & 37/50 & $2.5\times 10^4$ & $4.1\times 10^3$ & 30/50 & $3.1\times 10^4$ & $1.0\times 10^4$\\
\hline
\end{tabular}
\end{center}
\caption{Number of injections detected, mean and $\sigma$ for the distribution 
of the likelihood values for the GW and non-GW injections}
\label{thlike}
\end{table}

\subsubsection{Likelihood + determinant}
~\\
We now complete the maximum likelihood method with a cross check in
order to get rid of spurious events: this is achieved by means of the
geometrical method described in subsection \ref{subsdete}.

The major effect of this cross check is that in some cases
the value of the parameter $r$ defined in eq.(\ref{rdef}) is
above the threshold that we had chosen, $r_{thr}=0.05$,
which implies that a direction reconstruction through the linear algebra method
is simply not meaningful.
When this happens (in most of the non-GW injections and in a few GW ones),
the trigger is discarded.

\begin{figure}
\centering
\includegraphics[width=.6\linewidth,angle=90]{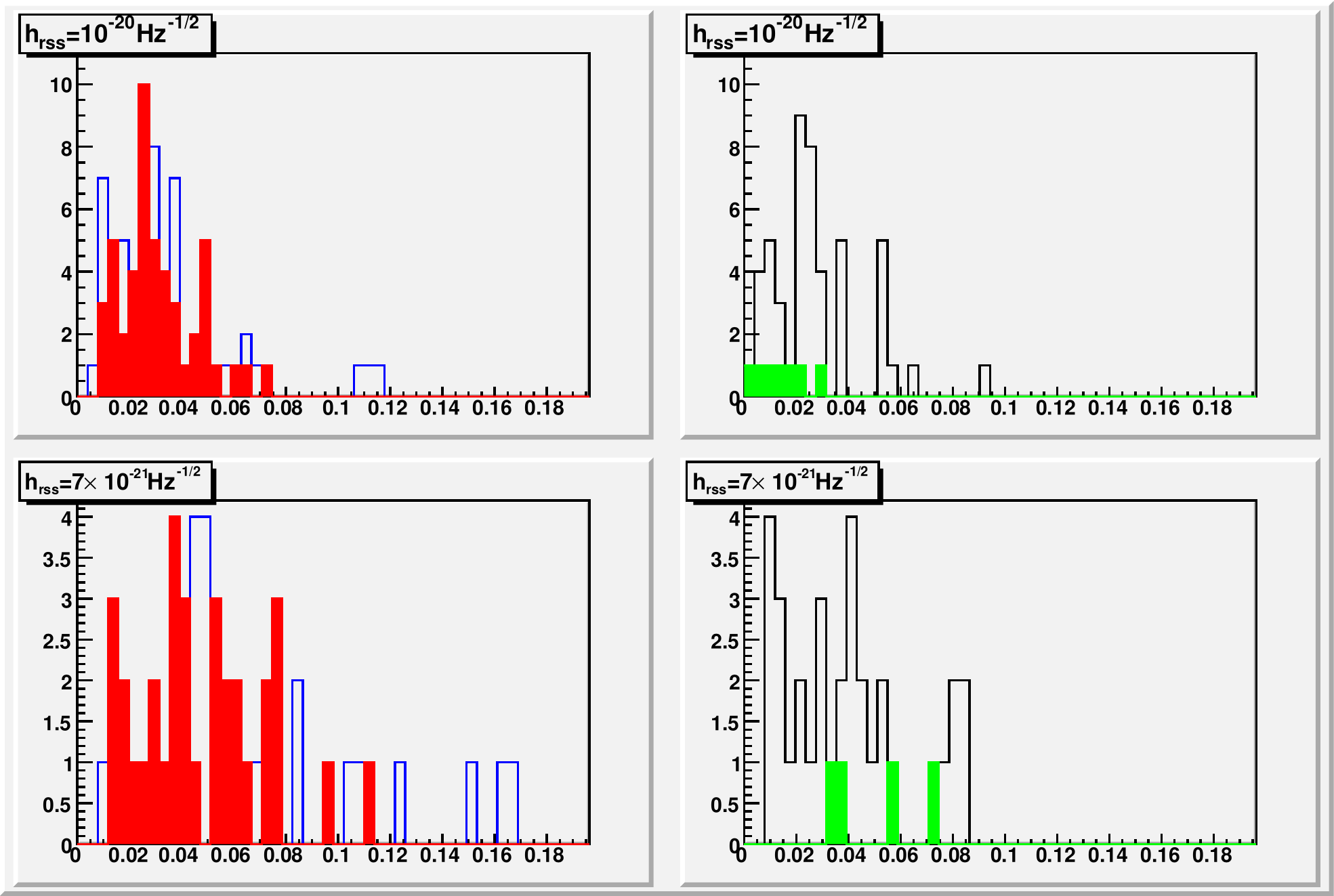}
\caption{Distribution of the distances between the injection direction and 
the direction reconstructed by the likelihood method (left, filled-red),
injection direction and the one reconstructed by the determinant method
(left, empty-blue), 
and between likelihood and determinant reconstructed directions
(right,empty-black),
for $h_{rss}=10$ (up) and $7$ (down) $\cdot 10^{-21}{\rm Hz}^{-1/2}$.
Also is displayed the distance in the direction determination for non
GW-injections (right, filled-green).}
\label{lag_1}
\centering
\includegraphics[width=.6\linewidth,angle=90]{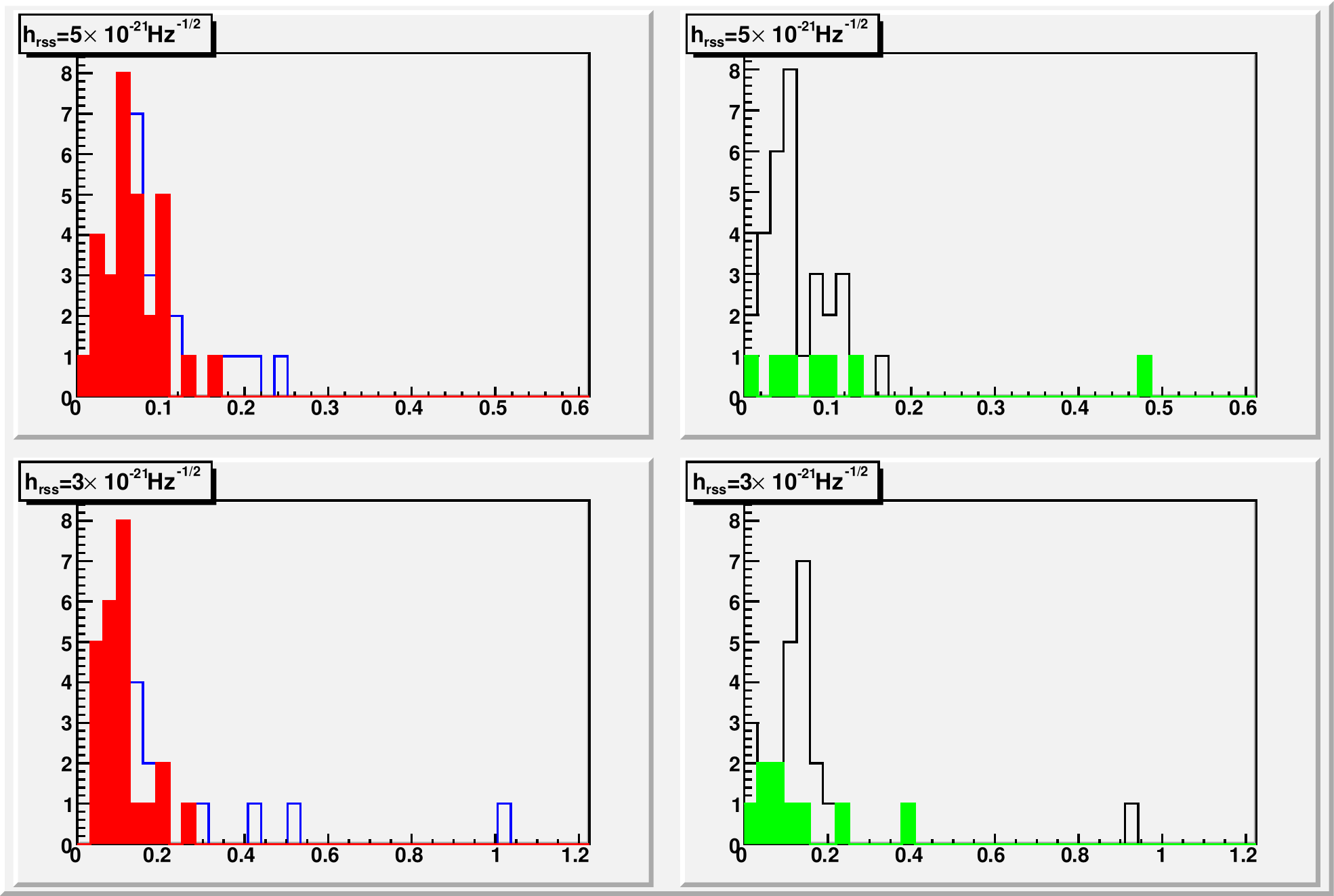}
\caption{Same as Fig.~\ref{lag_1} for 
$h_{rss}=5$ (up) and $3$ (down) $\cdot 10^{-21}{\rm Hz}^{-1/2}$.}
\label{lag_2}
\end{figure}

\begin{figure}
\centering
\includegraphics[width=.65\linewidth,angle=90]{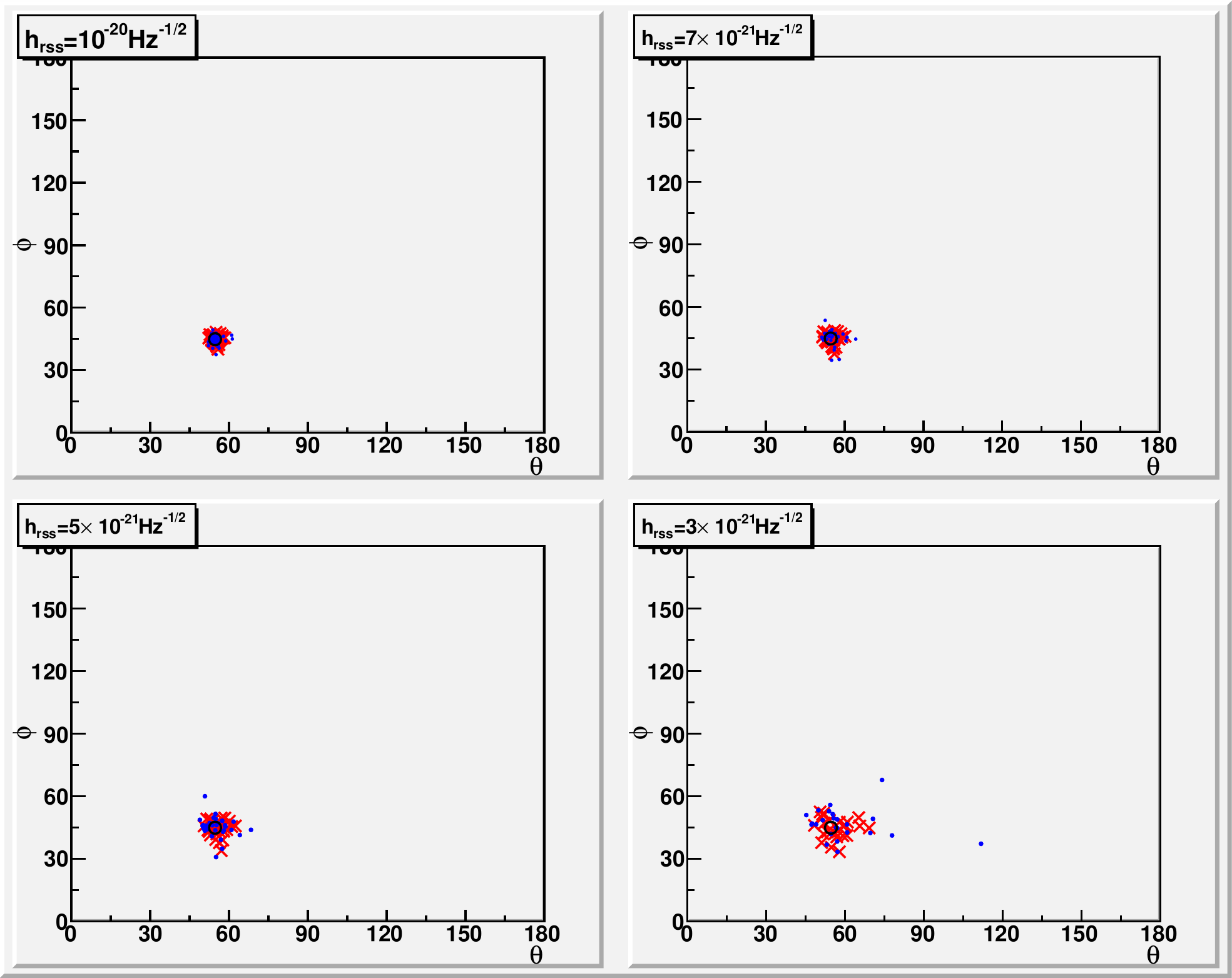}
\caption{Each graph correspond to 50 linearly polarized injections of a signal
(25$+$, 25$\times$).
The black circle marks the injection direction, red crosses indicate direction
reconstructed with the likelihood method, blue dots are obtained via the
determinant method.}
\label{sky_inj}
\end{figure}

In Figs.~\ref{lag_1},\ref{lag_2} we report the distribution 
of distances between different direction reconstructions,
see also tab.~\ref{tlag}, while fig.~\ref{sky_inj}
shows graphically the result of the direction reconstruction with the different
methods for the GW injections.
As expected (see \cite{Zhou:1995en}, \cite{Stevenson:1996rw}),
the precision in the determination of the arrival direction degrades
at decreasing SNR.

\begin{table}
\begin{center}
\begin{tabular}{|c|c|c|c|c|c|c|c|}
\hline
& \multicolumn{7}{|c|}{GW}\\
\cline{2-8}
$h_{rss}(10^{-21}{\rm Hz^{-1/2}})$ & & \multicolumn{2}{|c|}{i-l} &
\multicolumn{2}{|c|}{i-d} & \multicolumn{2}{|c|}{l-d} \\
\cline{3-8}
& \#$_{det}$/\#$_{trigg}$ & $\mu$ & $\sigma$ & $\mu$ & $\sigma$ & $\mu$ & $\sigma$ \\
\hline
10 & 48/50 & 0.031 & 0.015 & 0.037 & 0.026 & 0.027 & 0.018\\
\hline
7 & 33/48 & 0.048 & 0.024 & 0.062 & 0.040 & 0.040 & 0.023\\
\hline
5 & 30/45 & 0.068 & 0.035 & 0.089 & 0.055 & 0.063 & 0.036\\
\hline
3 & 24/37 & 0.11 & 0.053 & 0.19 & 0.21 & 0.15 & 0.18\\
\hline
\end{tabular}
\caption{Number, mean and $\sigma$ for the distributions of distances between 
injection direction and likelihood reconstructed (i-l), injection and 
determinant reconstructed (i-d), likelihood and direction (l-d) for GW 
injections for different signal strengths.}

\vskip 1pt

\begin{tabular}{|c|c|c|c|}
\hline
&\multicolumn{3}{|c|}{non-GW}\\
\cline{2-4}
$h_{rss}(10^{-21}{\rm Hz^{-1/2}})$ &\#$_{det}$/\#$_{trigg}$ & $\mu$ & 
$\sigma$\\
\hline
10 & 8/41 & 0.014 & 0.0089 \\
\hline
7 & 5/33 & 0.050 & 0.016\\
\hline
5 & 7/34 & 0.13 & 0.14\\
\hline
3 & 9/30 & 0.13 & 0.11\\
\hline
\end{tabular}
\caption{Number, mean and $\sigma$ for the distributions of distances between 
likelihood and determinant reconstructed directions for non-GW injections for 
different signal strengths.}
\label{tlag}
\end{center}
\end{table}

Finally, the application of the cut to eliminate triggers with
$\delta>\delta_{thr}$ does not change much the situation at this stage:
we found indeed that when a direction reconstruction is possible,
this is almost always compatible with the one found with the
likelihood algorithm.

\begin{figure}
\begin{center}
\includegraphics[width=.6\linewidth]{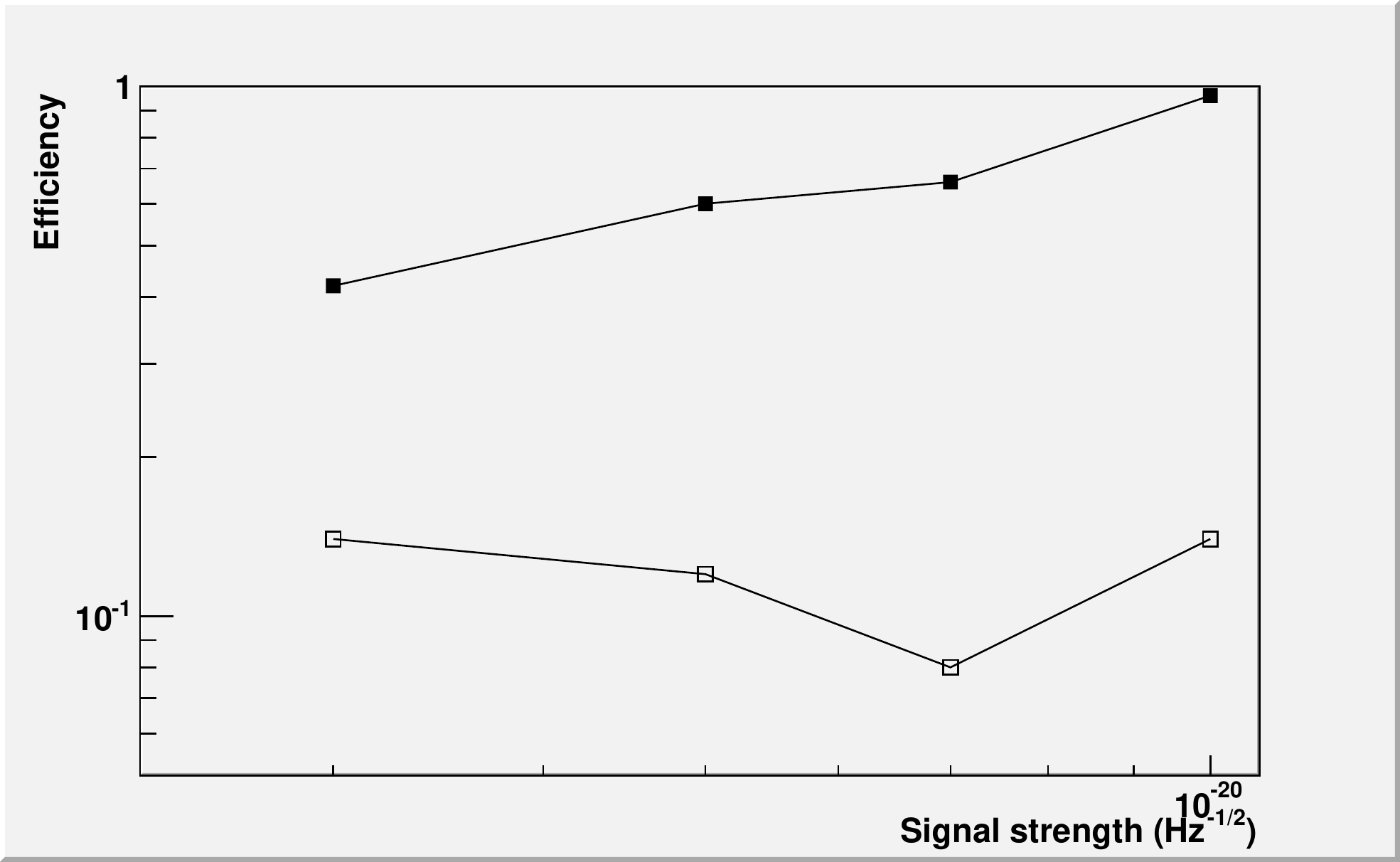}
\caption{Detection efficiency of GW-injections (filled squares) and of non-GW
signals (empty squares) as a function of the injected signal strength.
The triggers surviving all the cuts are (in order of decreasing SNR)
48,33,30 and 21 for the GW injections, and 7,4,6 and 8 for non-GW.}
\label{roc}
\end{center}
\end{figure}

Figure \ref{roc} shows the final detection efficiency for all the sets
of injections: it should be noted that the efficiency for non-GW seems
not to depend on the signal SNR, which points towards the fact that
they ''survived'' the cuts simply because they had by chance a GW-like
geometrical configuration. 
  
\section{Conclusions}
We have set up an algorithm which exploits the multichannel
capability of the sphere in such a way to filter spurious disturbances.
We have tested our method on various sets of injections of GW and
non-GW signals over a Gaussian background noise which reproduces a
possible outcome of the six transducers of a spherical detector like
miniGRAIL.

We found that by crossing the two direction reconstruction methods,
the false alarm rate due to Gaussian noise can be reduced to
${\cal O}(10^{-2}\div 10^{-3}{\rm Hz}^{-1})$ and that the detection efficiency
for non-Gaussian noise (the non-GW injections) is $5-10$\%, independently of
the SNR.
These results can be achieved while keeping the detection efficiency for
GW signals in the range $50-95$\%.

Moreover as summarized in tab.~\ref{tlag}, the arrival direction of the
gravitational wave can be determined with good accuracy on the trigger
satisfying our cross-check, with a precision which ranges from roughly $0.2$ 
rad ($\simeq 10^o$) at SNR$\simeq16$ to rapidly improve with the signal
strength up to $0.03$ ($\simeq 2^o$) at SNR$\simeq 53$.

\section*{Acknowledgments}
The authors acknowledge support by the Boninchi foundation and the Fonds 
National Suisse; R.S. acknowledges the support by an INFN grant during most of 
the completion of this work. 
We would like to thank Giorgio Frossati and Sasha Usenko
for giving us informations about the miniGRAIL configuration, Sergei 
Klimenko for kindly providing us the late version of the coherent WaveBurst, 
Gabriele Vedovato for his friendly help with WaveBurst, Michele Maggiore 
for encouragement and support and one of the referees for helping us
clarifying some important aspects of the theoretical model.

\begin{center}
\bf{Appendix A: detector parameters}
\begin{table}[htdp]
\begin{tabular}{|l|c|c|}
\hline sphere mass & $M$ & 1149 Kg \\
\hline sphere radius & $R$ & 0.0325 m \\
\hline sphere temperature & $T$ & 0.02 K\\
\hline fundamental quadrupole modes frequencies &&\\
 ($Y^2_{20}$,$Y^2_{1c}$,$Y^2_{1s}$,$Y^2_{2c}$,$Y^2_{2s}$)& $\omega_{N=0,\dots,5}$
& 2879,2872,2850,2741,2738 Hz \\
\hline scalar mode frequency ($Y_{00}$)& $\omega_{N=5}$ & 5902 Hz \\
\hline first excited quadrupole modes frequencies &&\\
($Y^2_{0}$,$Y^2_{1c}$,$Y^2_{1s}$,$Y^2_{2c}$,$Y^2_{2s}$) & $\omega_{N=6,\dots,10}$
& 5521,5515,5472,5263,5257 Hz \\
\hline  vector modes frequencies &&\\
($Y^1_{0}$,$Y^1_{1c}$,$Y^1_{1s}$)& $\omega_{N=11,\dots,13}$ & 3900,3890,3861 Hz \\
\hline {\cal l}=3 modes frequencies &&\\
($Y^3_{0}$,$Y^3_{1c}$,$Y^3_{1s}$,$Y^3_{2c}$,$Y^3_{2s}$,$Y^3_{3c}$,$Y^3_{3s}$)
& $\omega_{N=14,\dots,20}$ & 4291,4280,4247,4085,4081,4085,4081 Hz \\
\hline {\cal l}=4 modes frequencies &&5505,5491,5449,5241,5235\\
($Y^4_{0}$,$Y^4_{1c}$,$Y^4_{1s}$,$Y^4_{2c}$,$Y^4_{2s}$,$Y^4_{3c}$,$Y^4_{3s}$,$Y^4_{4c}$,$Y^4_{4s}$)
& $\omega_{N=21,\dots,29}$ & 5241,5235,5241,5235 Hz \\
\hline sphere modes quality factors & $Q_{N=0\dots29}$ & $2\times 10^6$ \\
\hline fundamental quadrupole modes &&\\
radial displacement & $\alpha_{N=0\dots4}$ & -2.88911\\
\hline scalar mode radial displacement & $\alpha_{N=5}$ & -3.4098\\
\hline first excited quadrupole modes &&\\
radial displacement & $\alpha_{N=6\dots10}$ & 0.0744804\\
\hline vector modes radial displacement & $\alpha_{N=11\dots13}$ & 0.818002\\
\hline {\cal l}=3 modes radial displacement & $\alpha_{N=14\dots20}$ &- 3.34268\\
\hline {\cal l}=4 modes radial displacement & $\alpha_{N=21\dots29}$ &- 3.63078\\
\hline & $\chi $ & -0.3278\\
\hline & $\chi_5 $ & -3.8043\\
\hline
\end{tabular} \caption{Sphere features}
\label{defaulttable}
\end{table}
\end{center}

\begin{table}[htdp]
\begin{center}\begin{tabular}{|l|c|c|}
\hline transducers' azimuth & $\theta_{k=0,1,2}$ & $37.3773^\circ$\\
\hline transducers' azimuth & $\theta_{k=3,4,5}$ & $79.1876^\circ$\\
\hline transducers' longitude & $\varphi_{k=0,1,2}$ & $(1+2k)*60^\circ$ \\
\hline transducers' longitude & $\varphi_{k=3,4,5}$ & $2k*60^\circ$ \\
\hline transducers' frequencies & $\omega_{k=0,1}$ & 2863 Hz \\
\hline transducers' frequencies & $\omega_{k=2,3}$ & 2850 Hz \\
\hline transducers' frequencies & $\omega_{k=4,5}$ & 2878 Hz \\
\hline transducers' masses & $m_{k=0,1}$ & 0.205 Kg \\
\hline transducers' masses & $m_{k=2,3}$ & 0.153 Kg \\
\hline transducers' masses & $m_{k=4,5}$ & 0.150 Kg \\
\hline transducers' quality factors & $Q_{k=0\dots6}$ & $1\times 10^6$ \\
\hline gap electric fields & $E_{k=0\dots5}$ & $2\times 10^7 V/m$ \\
\hline effective resistances & $r_{k=0,1}$ & 0.0883 Ohm\\
\hline effective resistances & $r_{k=2,3}$ & 0.1140 Ohm\\
\hline effective resistances & $r_{k=4,5}$ & 0.0872 Ohm\\
\hline total capacities & $C_{k=0,1}$ & 1.1638$\times 10^{-9}$ F\\
\hline total capacities & $C_{k=2,3}$ & 0.6978$\times 10^{-9}$ F\\
\hline total capacities & $C_{k=0,1}$ & 1.1935$\times 10^{-9}$ F\\
\hline primary inductances & $L^p_{k=0\dots5}$ & 0.3595 H\\
\hline secondary inductances & $L^s_{k=0\dots5}$ & 2.1$\times 10^{-6}$ H\\
\hline auto-inductances & $M^{ps}_{k=0\dots5}$& 1$\times 10^{-8}$ H\\
\hline SQUID's input inductances & $L^i_{k=0\dots5}$ & 1.7$\times 10^{-6}$ H\\
\hline SQUID's self inductances & $L^{SQ}_{k=0\dots5}$ & 8$\times 10^{-11}$ H\\
\hline SQUID's mutual inductances & $M^{SQ}_{k=0\dots5}$ & 1$\times 10^{-10}$ H\\
\hline SQUID's shunt resistances & $R^{sh}_{k=0\dots5}$ & 20 $\Omega$ \\
\hline
\end{tabular} \caption{Readout specifications}
\end{center}
\label{defaulttable_II}
\end{table}

\end{document}